\documentclass{aa}  
\usepackage{natbib}
\bibpunct{(}{)}{;}{a}{}{,}
\usepackage{graphicx}
\usepackage[colorlinks=true,linkcolor=blue,urlcolor=blue,citecolor=blue]{hyperref}
\usepackage{txfonts}
\usepackage{mathtools}
\makeatletter
\renewcommand*\aa@pageof{, page \thepage{} of \pageref*{LastPage}}
\newcommand{\teff}{\ensuremath{T_{\mathrm{eff}}}}
\newcommand{\logg}{\ensuremath{\log g}}
\newcommand{\vt}{\ensuremath{\xi}}
\newcommand{\xh}[1]{\ensuremath{[\mathrm{#1}/\mathrm{H}]}}
\newcommand{\xfe}[1]{\ensuremath{[\mathrm{#1}/\mathrm{Fe}]}}
\newcommand{\kms}{\ensuremath{\mathrm{km\,s^{-1}}}}
\newcommand{\fig}[1]{Fig.~\ref{#1}}
\usepackage{lscape}
\makeatother
\usepackage{xcolor}
\usepackage{gensymb}
\usepackage{multirow}
\usepackage{placeins}
\usepackage[caption = false]{subfig}
\usepackage{rotating}

\begin{document}

   \title{Chemical characterisation of small substructures in the local stellar halo\thanks{The full Table \ref{table:linelist} and Table \ref{tab:abundances} are available in electronic form at the CDS via anonymous ftp to \url{cdsarc.u-strasbg.fr} (130.79.128.5) or via \url{http://cdsweb.u-strasbg.fr/cgi-bin/qcat?J/A+A/542/A52}}}
   \author{Emma Dodd\inst{1,2}\thanks{email: emma.l.dodd@durham.ac.uk}
   \and Tadafumi Matsuno\inst{3}
   \and Amina Helmi\inst{1}
   \and Eduardo Balbinot\inst{1}
   \and Thomas M. Callingham\inst{1} 
   \and Else Starkenburg \inst{1}
   \and Hanneke C. Woudenberg\inst{1}
    \and Tomás Ruiz-Lara\inst{4,5}
          }

   \institute{
      Kapteyn Astronomical Institute, University of Groningen, Landleven 12, 9747 AD Groningen, The Netherlands\relax
      \and Institute for Computational Cosmology, Department of Physics, Durham University, South Road, Durham DH1 3LE, UK\relax
      \and Astronomisches Rechen-Institut, Zentrum f\"{ u}r Astronomie der Universit\"{ a}t Heidelberg, M\"{ o}nchhofstr. 12-14, 69120 Heidelberg, Germany\relax
      \and Universidad de Granada, Departamento de Física Teórica y del Cosmos, Campus Fuente Nueva, Edificio Mecenas, E-18071, Granada, Spain\relax
       \and Instituto Carlos I de Física Te\'orica y computacional, Universidad de Granada, E-18071 Granada, Spain\relax
             }

   \date{}

  \abstract
   {The local stellar halo of the Milky Way is known to contain the debris from accreted dwarf galaxies and globular clusters in the form of stellar streams and over-densities in the space of orbital properties (e.g. integrals of motion).}
   {While several over-densities have been uncovered and characterised dynamically using\textit{ Gaia} data, their nature is not always clear. Especially for a complete understanding of the smaller halo substructures, the kinematic information from \textit{Gaia} needs to be coupled with chemical information.} 
   {In this work, we combine \textit{Gaia} data with targeted high-resolution UVES spectroscopy of five small substructures that were recently discovered in the local halo, namely ED-2, -3, -4, -5, and -6 (the ED streams). We present the chemical abundances measured from our newly obtained UVES spectra (20 stars) and from archival UVES spectra (nine stars). We compared these with homogeneously derived abundances from archive spectra of 12 \textit{Gaia} Enceladus (GE) stars.} 
   {The chemical abundances of all five substructures suggest that they are of accreted origin, except for two stars that present a high [$\alpha$/Fe] at high [Fe/H] more in line with an in situ origin. All but ED-2 present a significant spread in [Fe/H] suggestive of a dwarf galaxy origin. ED-3 and ED-4 tend to exhibit a lower [$\alpha$/Fe] compared to GE stars. 
   As for ED-5 and ED-6, they are consistent with the GE chemical track and could be high-energy tails of GE that were lost earlier in the accretion process. We present new elemental abundances for five ED-2 stars, including more elements for the \textit{Gaia} BH3 companion star. Our findings are in line with the picture that ED-2 is a disrupted ancient star cluster.}
   {}


   \maketitle
%

\section{Introduction}
The stellar halo of the Milky Way is full of stellar streams from accreted dwarf galaxies and disrupted globular cluster (GC) systems \citep[e.g. see recent reviews from][]{helmi2020,deason2024galactic,bonaca2024stellar}. As an accreted system is torn apart by tidal forces, its stars initially remain spatially and kinematically coherent in the form of a stream. These stars then phase-mix over time on a timescale that depends on the orbit and the internal properties of the system.
For example, streams from GCs typically remain coherent for long timescales due to their lower velocity dispersions. 
Many of these kinematically cold stellar streams have been discovered in the distant halo, detected as over-densities on the sky using deep photometric surveys \citep[starting from the pioneering SDSS:][see e.g. \citealt{belokurov2006field}]{York2000}. 

Since \textit{Gaia} \citep{prusti2016gaia} 
has provided positions, photometry, and proper motions across the full sky, there has been a surge in the number of stream detections. 
A large number of these discoveries have been made with the \verb|STREAMFINDER| algorithm. It was developed to look for streams in the full \textit{Gaia} data using all the observable information as input, and it is able to pick out stars that are on similar orbits and have similar stellar populations \citep{malhan2018streamfinder,ibata2024charting}. 
\verb|STREAMFINDER| has played an important role in identifying many cold stellar streams across the Galaxy, but it has limitations at larger distances and in the plane ($|b|<20 \degree$) due to computational resources and the increased density of (contaminating) stars. 

In the vicinity of the Sun, \textit{Gaia} provides full 6D phase-space information for stars, allowing for characterisation of halo streams in integrals of motion (IoM) space, which are properties that remain conserved along an orbit and with time \citep[e.g. energy and angular momentum in a time-independent axisymmetric potential,][]{helmi2000}.
Since stars in a stream move along similar orbits, they are also apparent as tight clumps in integrals of motion space, even long after being phase mixed. Their association with other streams or GCs can be investigated from their proximity in these spaces \citep{Bonaca:2021}. Notably, it has been predicted that the local halo contains hundreds of stellar streams that can be identified as kinematically coherent substructures \citep[e.g.][]{helmi1999b,gomez2013streams}.

Recently \textit{Gaia} DR3 data \citep{GaiaDR3_Summary} have revealed an increasing number of small substructures in the stellar halo near the Sun, including several small groups of (10-30) stars that are tightly clustered in IoM space \citep[e.g.][or velocity space e.g. \citealt{Mikkola:2023}]{Dodd2023,Tenachi_Typhoon_22_arXiv220610405T,Oria_Anaeus_22_2022arXiv220610404O}. Several of these streams have loosely bound orbits (with low binding energy) suggestive of an accreted origin. However, it is not known whether their progenitors were dwarf galaxies or disrupted GCs or if they can be related to already known debris from a large merger. 
Some of the substructures overlap with regions where one might expect debris from \textit{Gaia}-Enceladus (GE), although this depends on the exact configuration of the merger \citep[e.g. see][]{koppelman2020,Naidu2021,amarante2022}. 

To map out the debris from larger mergers and understand which halo substructures are distinct, properties to link together fragmented debris are needed.
This is difficult to do with dynamics alone; however, chemical abundances can be used in parallel. The chemical composition at the stellar surface remains largely unchanged during the evolution of a star, and therefore the chemical patterns in individual systems reflect their star formation and chemical enrichment histories. We now know that GE stars show a distinct chemical track, for example in [Fe/H]--[$\alpha$/Fe] space at high metallicity when compared with in situ stars \citep[e.g.][]{nissen2010,helmi_merger_2018}. Other accreted substructures such as the Helmi streams \citep{matsuno2022high} and Sequoia \citep{matsuno2022,Ceccarelli2024} present lower [$\alpha$/Fe] than GE at a fixed \xh{Fe}. High-precision and homogeneous abundances are essential to be able to detect such differences and disentangle whether the smaller substructures in the local halo are new accretion events or fragments of already known structures. For example, if they are high-energy tails of GE, we can use them to further characterise how the merger took place and determine whether it was prograde or retrograde and what was the inclination ~\citep[][]{koppelman2020}. 

However, chemically linking debris at high energy to already known structures is not straightforward, as one has to consider that any chemical gradients in the initial progenitor will be present in the orbital properties of the debris. For example, the more metal-poor debris in the outskirts of the progenitor is stripped at earlier times and therefore deposited onto higher energy orbits \citep[][]{amarante2022}. 
Although this complicates the matter, stars from the same progenitor should still follow the same chemical sequence; for example, the lower [Fe/H] stars in the outskirts will have a higher [$\alpha$/Fe]. Therefore, with detailed chemical abundances in many elements, the association of fragmented debris should still be possible.

The retrograde halo contains several smaller substructures within the Sequoia \citep{myeong2019,matsuno2019} region of IoM space, including Antaeus/L-RL-64 \citep{Oria_Anaeus_22_2022arXiv220610404O,ruizlara2022}, ED-2, ED-3 \citep{Dodd2023}, I'itoi, and Arjuna \citep[][]{naidu2020}. However, it has been suggested that Arjuna could be GE debris in the retrograde halo \citet{Naidu2021}. Uncovering the nature of these small retrograde substructures requires targeted chemical follow-up studies. One such study was conducted by \citet{Ceccarelli2024} with the A Walk on the Retrograde Side (WRS) project. Their aim was to provide chemical abundances derived homogeneously from high-resolution UVES spectra for a couple of hundred local retrograde halo stars. 
The first release presents the abundances for 186 stars, agreeing with \citet{matsuno2022} that Sequoia is lower in [$\alpha$/Fe] than GE at a fixed \xh{Fe} and suggesting a single progenitor for ED-3 and Antaeus/L-RL-64 based on their similar chemical patterns.

Some of the smaller substructures are expected to be disrupted GCs, and if proven to be so, they could provide important clues to the formation of GCs. Although there are no observed GCs below the metallicity floor (\xh{Fe} $\sim$ -2.4) in the Milky Way, several stellar streams are known to exhibit a lower metallicity with a small metallicity dispersion that suggests they are disrupted GCs.  For example,  the local retrograde ED-2 stream lies around the metallicity floor, with a \xh{Fe} of $-2.5$ dex \citep{balbinot2023,balbinot2024}; the Phoenix stream has \xh{Fe} of $-2.7$~dex \citep{Balbinot:2016,Wan:2020} and C19 even lower with  \xh{Fe} of $-3.4$~dex \citep{Ibata:2021,Martin:2022}.

Understanding the origin of such streams is critical to constraining the formation and disruption of low-metallicity GCs and identifying whether there is a preferential disruption of metal-poor GCs such that they are more likely to be found as disrupted streams \citep{Kruijssen:2019}.
Moreover, one of the streams, ED-2, is now known to host the Gaia-BH3 33 $M_\odot$ black hole \citep{Panuzzo:2024,balbinot2024}. Understanding the properties of ED-2's progenitor cluster can provide useful insight into the origin of such massive stellar black holes, which is still challenging to explain \citep{abbott2023population}.

This work aims to characterise several of the small substructures discovered in the local stellar halo by combining the dynamical information from \textit{Gaia} with chemical abundance information from spectroscopy.
The tight clumps in IoM space that were discovered in \citet{Dodd2023}, namely, ED-2, -3, -4, -5, and -6 are the topic of this paper. We provide follow-up for these streams with high-resolution spectroscopy using UVES, from which we derived homogeneous and high-precision chemical abundances that allowed us to investigate the nature of their progenitors. 

This paper is structured as follows. We present the data and the chosen targets in Section \ref{sec:data}, including the archival spectra that supplement this dataset. Then we present the chemical abundance analysis in Section \ref{sec:abundances} and its results in Section \ref{sec:results}, including measurements of the metallicities and dispersions in Section \ref{sec:metallicities} and the abundances of C, N, and odd-Z elements; $\alpha$ elements; iron peak elements; and neutron capture elements. Based on these results, we present our interpretation of the nature of the ED streams throughout Section \ref{sec:discussion}. There, we also discuss the accreted or in situ origin of the streams and the nature of their progenitors. 
We present our conclusions in Section \ref{sec:conclusion}.

\section{Data} \label{sec:data}

We obtained high-resolution spectra for 21 stars in the small high-energy groups identified in \citet{Dodd2023} using the Ultraviolet and Visual Echelle Spectrograph (UVES) \citep{Dekker:2000} on the Very Large Telescope (VLT) of the European Southern Observatory (111.D-0263A PI Dodd). This includes five stars in ED-2, five stars in ED-3, four stars in ED-4, three stars in ED-5, and five stars in ED-6.
One of ED-2 stars, 5991844282681283712, is a RR Lyrae type-c star for which we do not derive chemical abundances. However, we note that in the \textit{Gaia} DR3 RR Lyrae catalogue, \citep{Clementini2023} has a photometric [Fe/H] of $-2.18 \pm 0.24$~dex, 
and the updated \citealt{li2023photometric} catalogue\footnote{This catalogue has a newly calibrated relation between pulsation period and [Fe/H].} reports a photometric [Fe/H] of $-2.46 \pm 0.20$~dex. 
The chemical abundances for ED-2 stars were already presented in \citet{balbinot2024}, although here we also provide abundances for all the other measured elements. 

\begin{figure}
\centering
\includegraphics[width=\linewidth]{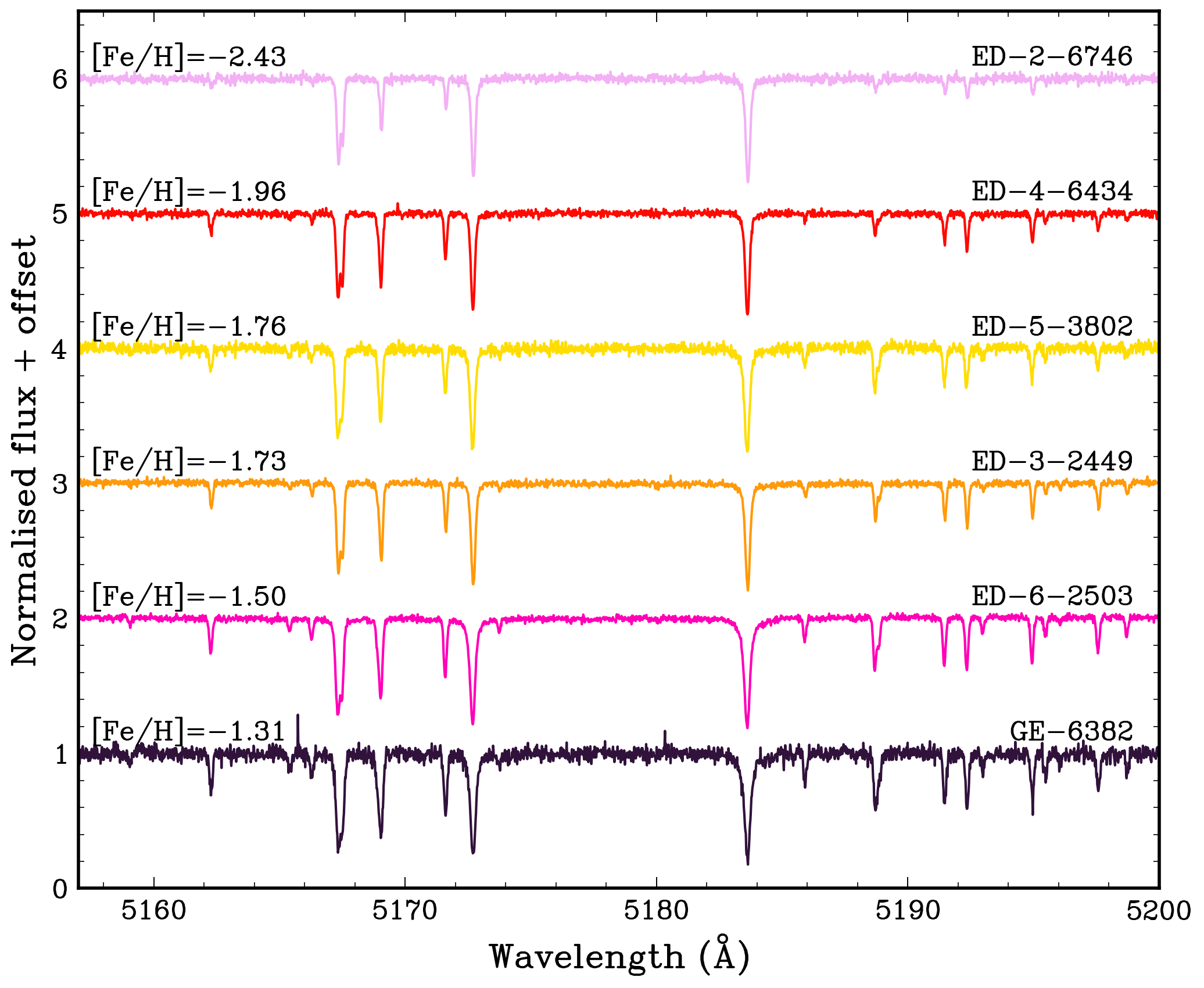}
\caption{Spectra around the Mg triplet region of one star in each ED substructure and one GE star. The stars have been selected to have a similar $T_{eff}$ within the range of 6150\,K -- 6450\,K and span a range of \xh{Fe}, demonstrating the difference in the lines for more metal-poor stars. The GE spectrum is plotted such that the continuum is normalised to 1 and then each other spectra are plotted with an offset of 1, ordered from most metal-rich to most metal-poor. 
} 
\label{fig:spectra}
\end{figure}

Additionally, we searched for archive UVES data of stars in these groups and found two ED-2 stars \citep[including the \textit{Gaia} BH3 companion star 4318465066420528000,][shown to be associated with ED-2 in \citealt{balbinot2024}]{Panuzzo:2024}, four ED-3 stars, two ED-4 stars and one ED-5 star.
We further supplement these datasets with archival data for 12 \textit{Gaia} Enceladus (GE) stars for comparison, selected according to the definition of GE in \citet{Dodd2023}. 
We used Phase 3 data products for further analyses, and the details of the observations and reduced data are given in Table~\ref{tab:obs}. Example spectra, around the Mg triplet region, of one star in each of the ED streams and one GE star for comparison are provided in Fig.~\ref{fig:spectra}.

\subsection{Targets}\label{sec:targets}
The targets were chosen following the definitions for the local halo substructures identified in \citet{Dodd2023}, by applying a data-driven and statistically robust clustering algorithm \citep{lovdal2022,ruizlara2022} on the \textit{Gaia} DR3 RVS sample. 
Here we repeat the relevant information of the creation of the local halo sample used in \citet{Dodd2023}.

\begin{figure*}
\centering
\includegraphics[width=\textwidth]{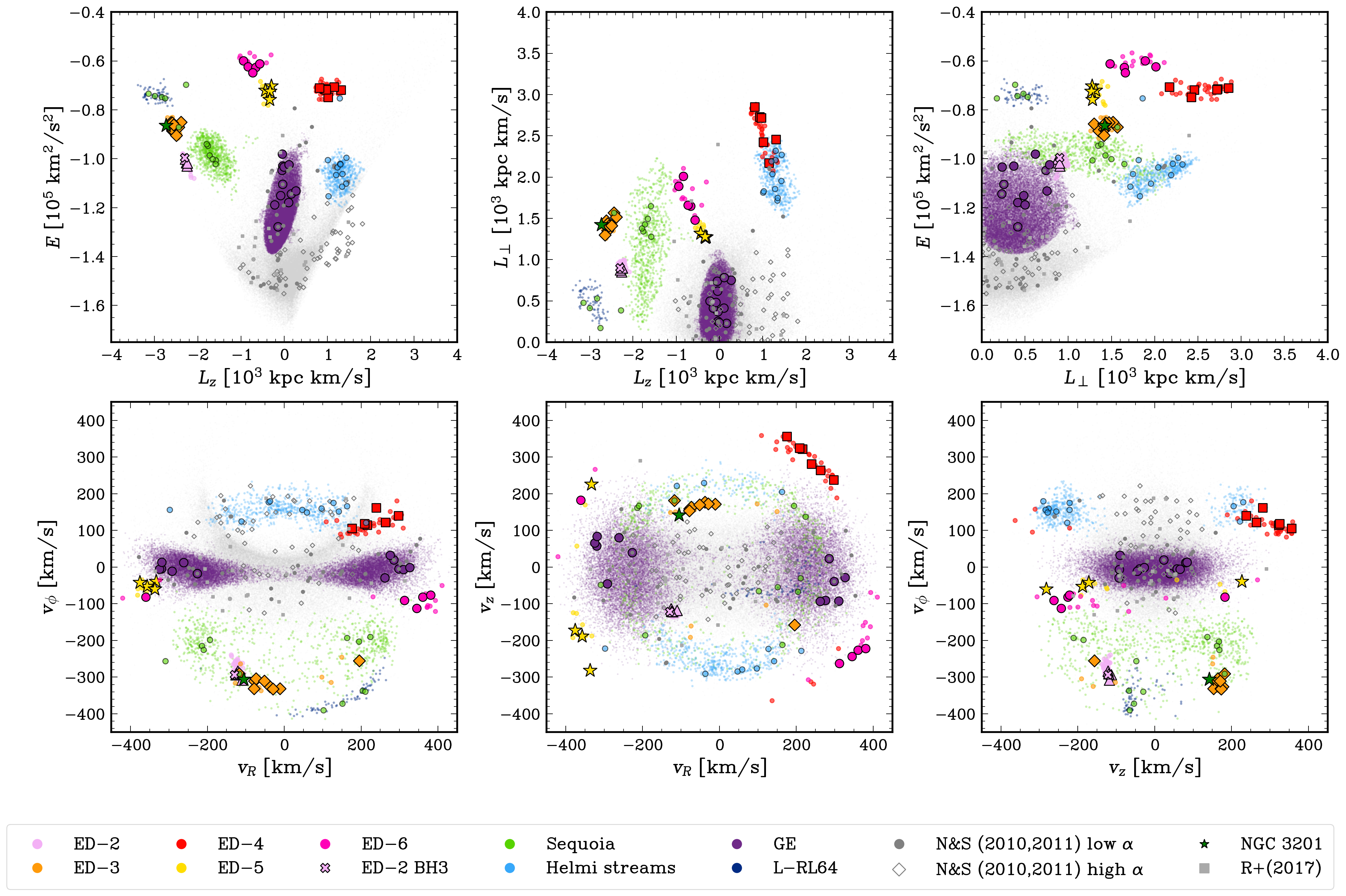}
\caption{Integrals of motion (top row) and velocity (bottom row) space for local halo stars, shown in light grey. The substructures relevant to this work are shown by coloured markers, and those that we have derived chemical abundances of are marked as larger symbols with a black outline. The \citet{nissen2010,nissen2011} and \citet{reggiani2017} samples are included here since we used these datasets as a comparison. For  \citet{nissen2010,nissen2011} stars the sample is split into high-$\alpha$ and low-$\alpha$ samples, shown as grey open diamond symbols and grey filled circles respectively. The \citet{reggiani2017} sample are shown as grey open square symbols. The Helmi Streams and Sequoia stars with abundances from \citet{matsuno2022,matsuno2022high} are shown in the relevant colour with a black outline. We also show the position of NGC 3201 as a green star since it is located very close to the ED-3 stars in all three IoM. } 
\label{fig:IOM}
\end{figure*}

\begin{figure*}
\centering
\includegraphics[width=0.95\textwidth]{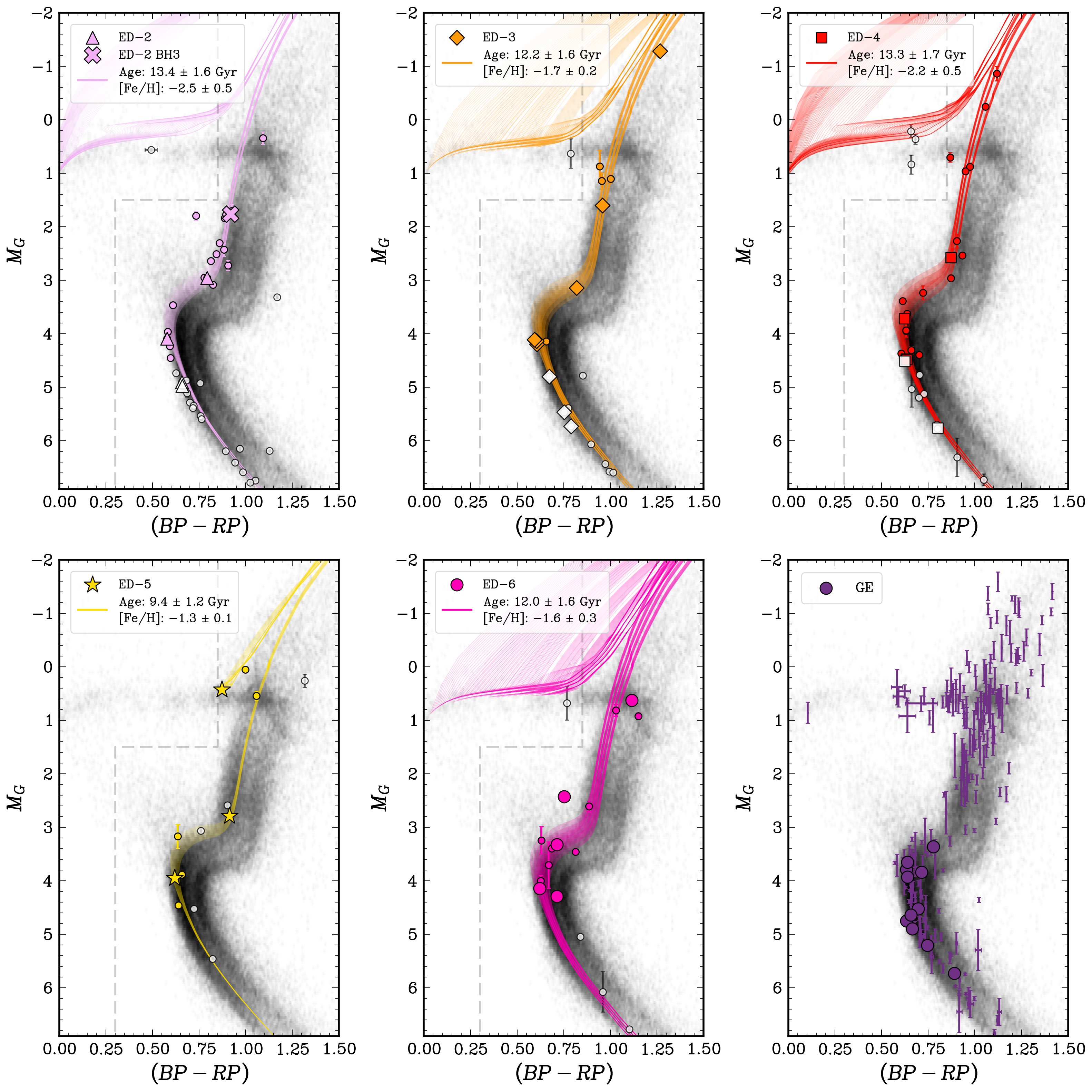}
\caption{Colour magnitude diagrams of ED groups. The stars for which we measure chemical abundances are shown as larger markers with different symbols. The last panel shows a randomly selected 1\% sample of GE stars and those for which we measure chemical abundances are shown with larger markers. In all cases, the colours and magnitudes have been corrected for extinction using the \citet{L22} dust maps. Horizontal branch stars have been removed from the isochrone fitting of the ED groups, indicated by the grey dashed line. Stars with high extinction $A_G > 0.7$ or that are fainter than M$_G$ = 4.5 are also removed from the isochrone fitting. All of these stars not included in the isochrone fitting are shown as white symbols. The compatible BaSTi IAC alpha-enhanced isochrones for each of the ED groups are shown and the range of age and \xh{Fe} that these isochrones span are given in the legend. The colour-magnitude shift ($\Delta_c$, $\Delta_m$) of ($-0.035$,  $0.04$) from \citet{gallart2024} has been applied to the isochrones for fitting and plotting. 
The background shows a density plot of the local halo stars.
} 
\label{fig:CMDs}
\end{figure*}

Stars were selected with (total) relative parallax uncertainty less than 20\% as $(\varpi -\Delta_\varpi) /\sqrt{\sigma_{\textrm{parallax}}^2 + \sigma_{\textrm{sys}}^2} \ge 5 $,
where $\Delta_\varpi$ represents the individual zero-point offsets determined following \citet{lindegren2021}, $\sigma_{\textrm{parallax}}$ is the \texttt{parallax\_error}, 
and $\sigma_{\textrm{sys}}$ is the systematic uncertainty on the zero-point, which was taken to be 0.015 mas  \citep{lindegren2021}. This corrected parallax was inverted to obtain distances to all stars and those within 2.5\,kpc of the Sun were selected. 
The following quality cuts were applied to the sample: \texttt{RUWE} $< 1.4$, line of sight velocity uncertainty\footnote{after applying the correction to \texttt{radial\_velocity\_error} recommended by  \citet{Babusiaux2022}} $\epsilon\,(V_{\rm los}) < 20$~\kms, \texttt{(G$_{RVS}$ - G)}$> -3$, and stars with $|b| < 7.5$ required
\verb|rv_expected_sig_to_noise|~$>5$.
The solar motion was corrected for using 
$ (U, V, W )_\odot$ = (11.1, 12.24, 7.25) \kms \citep{schonrich2010}
and the motion of the local standard of rest (LSR) using a $|\textbf{V}_{\textrm{LSR}}|$ of 232.8\,\kms \citep{mcmillan2017}. Halo stars were selected as |$\textbf{V}-\textbf{V}_\textrm{LSR}$| $>$ 210 \kms, resulting in a 6D halo sample of 69,106 objects. This is the sample to which the clustering algorithm was applied and how the ED substructures were originally identified. 
Integrals of motion (IoM) for stars (energy; $E$, and angular momenta; $L_z$, $L_{\perp}$) were calculated assuming $R_{\odot}$ = 8.2\,kpc \citep{mcmillan2017} 
and a Galactic potential which consists of a Miyamoto-Nagai disc with parameters $(a_d, b_d) = (6.5, 0.26)$~kpc, $M_{d}=9.3\times 10^{10} M_\odot$,
a Hernquist bulge with $c_b = 0.7$~kpc, $M_{b}=3.0 \times 10^{10}~M_\odot$, 
and an NFW halo with  $r_s=21.5$~kpc, $c_h$=12, and $M_{\rm halo}=10^{12}~M_\odot$. The sign of $L_z$ is flipped to be positive for prograde stars. This is the catalogue provided in \citet{Dodd2023} and used in this paper.

When searching for archival data we extended the sample from only \textit{Gaia} DR3 RVS to add extra stars with radial velocities from ground-based spectroscopic surveys, namely LAMOST DR7 \citep{cui2012}, APOGEE DR17 \citep{accetta2022}, GALAH DR3
\citep{Buder2020}, and RAVE DR6 \citep{steinmetz2020}.
We add these extra stars to our predefined substructures, by selecting those that fall within an ellipsoid in IoM space ($E$, $L_z$, $L_{\perp}$) which contains 80\% of the original substructure members. This choice of size for the ellipsoids was found previously to minimise adding noise to substructures (see \citet{ruizlara2022} for more details).

Figure~\ref{fig:IOM} shows this 6D halo sample within 2.5~kpc, with the substructures relevant to this work highlighted for comparison, including \textit{Gaia} Enceladus, Sequoia, the Helmi streams and L-RL64/Antaeus. Stars with spectra and for which we have derived chemical abundance information are shown as markers with black outlines within ED-2,-3,-4,-5,-6 and GE. We also show here the \citet{nissen2010,nissen2011}, and \citet{reggiani2017} samples of general halo stars which we use as a comparison for our abundances. 

From our kinematically defined groups, we observe stars selected to have $M_G<6$ to avoid lower main-sequence stars. For each group, we observe the brightest stars that are visible from the southern sky.
Fig.~\ref{fig:CMDs} shows the CMDs of the different ED substructures with stars that we have obtained spectra for, as well as those taken from the archive, both shown with larger symbols. We have corrected for extinction using the \citet{L22} dust maps and using the transformation from E(B-V) to \textit{Gaia} extinction 
from \citet{fitzpatrick2019analysis}. 
\noindent In the last panel of Fig.~\ref{fig:CMDs}, we also show the CMD of GE stars including those with archival spectra with larger symbols. For this subset of GE stars, most are main sequence stars and there is one sub-giant branch (SGB) star.
There are some obvious outliers in the CMDs of ED-2 and ED-5 that are significantly redder than the other stars in the substructure, we note that these stars have a high extinction with $A_G >0.7$. In all panels of Fig.~\ref{fig:CMDs} the full halo sample is shown in greyscale in the background, for comparison with the red and blue halo sequences \citep{GaiaCollaboration2018b}. 

Table \ref{tab:obs} presents the stars for which we have obtained new spectra or for which we have compiled archive spectra. The observation date, proposal IDs and radial velocities measured from the spectra are presented here. Stars in the ED groups have signal-to-noise ratios in the red and blue of $\sim 30\,-\,166$ and $\sim 9\,-\,86$, respectively.

\subsection{Stellar populations}\label{sec:stellar_pops}
Along with the CMDs in Fig.~\ref{fig:CMDs}, we also show the best-fit BaSTI-IAC alpha-enhanced isochrones \citep{pietrinferni2021updated}. We applied a colour and magnitude shift ($\Delta_c, \Delta_m$) to the BaSTI-IAC isochrones, where $\Delta_c$, the shift in ($G_{\rm BP} -G_{\rm RP}$), is taken to be $-0.035$ 
and $ \Delta_m$, the shift in $M_G$, is taken to be $0.04$. Small systematic shifts between models and data in different bands are expected to exist and were determined for the \textit{Gaia} photometric system by \citet{gallart2024} from averaging the residuals of colour-magnitude fitting using synthetic CMDs consisting of BaSTI-IAC isochrones, applied to the high-quality \textit{Gaia} Catalogue of Nearby stars \citep{smart2021gaia}.

To determine the best-fit isochrones, we only used stars with a low reddening, $A_G <0.7$, that are intrinsically bright, $M_G<4.5$, and 
that have a good \texttt{phot\_bp\_rp\_excess\_factor} using
$$
 0.001+0.039 \times \texttt{bp\_rp} < \log(\texttt{phot\_bp\_rp\_excess\_factor})
$$
 and
$$ \log(\texttt{phot\_bp\_rp\_excess\_factor}) < 0.12 + 0.039\times \texttt{bp\_rp}, 
$$
where \texttt{bp\_rp} is the apparent $(G_{\rm BP}-G_{\rm RP})$ colour. We also only used stars redwards of the boundary shown by the grey dashed line in Fig.~\ref{fig:CMDs}, that is, stars that have colours $(G_{\rm BP}-G_{\rm RP}) > 0.3$, and for brighter stars with $M_G$>1.5, the colours have $(G_{\rm BP}-G_{\rm RP}) > 0.85$. This removed horizontal branch stars from the isochrone fitting.

To estimate which isochrones best fit the distribution of stars in the CMD, we measured the distance to each isochrone as
\begin{equation*}
 d_2 = \sqrt{\frac{\sum_{i=1}^{n}{|X_i-X_{i, iso}|^2}\times w_i}{\sum_{i=1}^{n}{w_i}}}, 
\end{equation*}
where ${|X_i-X_{i, iso}|^2}$ represents the distance of each star $i$ to the closest point of the isochrone and $w_i$ is the weight given to each star, defined as the inverse of the sum in quadrature of the photometric errors. 
We fit each group with a set of isochrones that cover a wide range of ages from 5.0 to 18.0 Gyr in steps of 0.1 Gyr and metallicities [Fe/H] from -3.3 to 0.1~dex, in unequal step width that is larger at lower metallicity (see Table 2 of \citealt{pietrinferni2021updated}). 
We took the isochrones that are consistent with the mean and within one standard deviation of all the isochrones in the lowest fifth percentile of the distribution of distances ($d_2$), and we show these in Fig.~\ref{fig:CMDs}. Since the grid is unequally spaced, this resulted in a larger uncertainty at lower metallicities, thus reflecting the coarser grid in the model isochrones.

From Fig.~\ref{fig:CMDs} we observed that ED-2 presents a very tight CMD that is bluewards of the blue halo sequence, in line with the low metallicity and small spread seen in \citet{balbinot2024}. Since in ED-2 there are not many stars around the MSTO and SGB, which are the most sensitive to age, there is a large spread in the possible isochrones that could fit due to the age-metallicity degeneracy. 
Our fits also suggest that ED-2 is the oldest of the ED streams, with an age estimate on the order of the age of the Universe, as also shown in \citet[][]{balbinot2024}, ED-2 is comparable in age or even older than M92, one of the oldest Galactic GCs. ED-3 and ED-4 also present a fairly tight sequence, although less narrow than ED-2. Their stars are also populating the blue sequence and mostly the bluer end of this. It is difficult to reproduce the SGB of ED-4 and also the spread in $M_G$ around the horizontal branch, which could hint at a superposition of populations, although we are limited in the number of stars. In ED-3 also, we are lacking stars around the MSTO and SGB, making it difficult to constrain the age.

Unlike ED-2, -3, and -4, the member stars of ED-5 and ED-6 are more spread out in the CMD and do not follow a narrow sequence, suggesting they are not single-age populations with a small metallicity spread. A single isochrone is not a good representation of these systems. Both ED-5 and -6 member stars appear to be mostly on the blue sequences with one or two red sequence stars. Especially for ED-6, we observed that a large range of isochrones fit the observations. For ED-5, the star in the red sequence at $M_G \sim 0.5$ has a high $A_G$ of 0.93 and the stars populating the fainter end of the MS are fainter than the $M_G<4.5$ cut and hence are not included in our isochrone fitting. As a result, the range of compatible isochrones is narrower. Considering this, the isochrones shown in Fig.~\ref{fig:CMDs} give an average age and metallicity for each ED and some indication of the spread in these properties, which we nvestigate further with high-precision chemical abundances.

\begin{table*}
    \centering
     \caption{First six lines of the line list table.}
    \begin{tabular}{cllllllll}
    \hline \hline
        \textit{Gaia} ID& Species& $\lambda$ ($\AA$) & $\chi$ (eV) & log($gf$) (dex) & EW (m$\AA$) & A($X$)$_{\rm{LTE}}$ (dex) & A($X$)$_{\rm{NLTE}}$ (dex) \\ \hline
        \hline
        4245522468554091904 & LiI & 6707.81 & 0.00 & 0.17 & 22.0 & 2.36 & ~ \\ 
        4245522468554091904 & NaI & 5889.95 & 0.00 & 0.11 & 75.8 & 3.66 & 3.44 \\
        4245522468554091904 & MgI & 5172.68 & 2.71 & -0.36 & 119.9 & 5.32 & 5.42 \\
        4245522468554091904 & MgI & 5183.60 & 2.72 & -0.17 & 136.2 & 5.35 & 5.41 \\
        4245522468554091904 & MgI & 5528.40 & 4.35 & -0.55 & 24.8 & 5.37 & 5.43 \\ 
        4245522468554091904 & AlI & 3944.01 & 0.00 & -0.63 & 46.1 & 3.28 & 3.72 \\  \hline
    \end{tabular}
    \label{table:linelist}
    \tablefoot{The full version of this table is available online at the CDS.}
\end{table*}

\begin{table*}
   \centering
   \caption{Abundances table for one star in the sample (ED-2-4245). 
   }
 
		\begin{tabular}{cllllll}
  \hline
  \hline 
			\textit{Gaia} ID& Species&$N$ &$\xh{X}$ (dex) &$\sigma_{\xh{X}}$ (dex) & $\xfe{X}$ (dex) &$\sigma_{\xfe{X}}$ (dex) \\
			\hline
             \hline
			4245522468554091904 & LiI   & 1 &  1.31 & 0.12 &  3.87 & 0.10 \\
			4245522468554091904 & NaI   & 1 & -2.80 & 0.12 & -0.03 & 0.10 \\
			4245522468554091904 & MgI   & 3 & -2.18 & 0.09 &  0.30 & 0.06 \\
			4245522468554091904 & AlI   & 2 & -2.78 & 0.10 & -0.70 & 0.07 \\
			4245522468554091904 & SiI   & 1 & -2.45 & 0.13 &  0.11 & 0.10 \\
			4245522468554091904 & CaI  & 9 & -2.14 & 0.06 &  0.41 & 0.03 \\
			4245522468554091904 & ScII & 1 & -2.33 & 0.11 &  0.14 & 0.14 \\
			4245522468554091904 & TiI  & 1 & -1.85 & 0.12 &  0.70 & 0.10 \\
			4245522468554091904 & TiII & 7 & -2.04 & 0.05 &  0.43 & 0.10 \\
			4245522468554091904 & CrI  & 3 & -2.67 & 0.10 & -0.12 & 0.06 \\
			4245522468554091904 & MnI   & 3 & -3.04 & 0.10 & -0.49 & 0.06 \\
			4245522468554091904 & FeI  & 40& -2.55 & 0.07 &  0.00 & 0.02 \\
			4245522468554091904 & FeII  & 4 & -2.47 & 0.10 &  0.00 & 0.14 \\
			4245522468554091904 & CoI  & 1 & -2.20 & 0.13 &  0.36 & 0.10 \\
			4245522468554091904 & NiI  & 1 & -2.46 & 0.12 &  0.09 & 0.10 \\
			4245522468554091904 & SrII   & 2 & -2.56 & 0.11 & -0.09 & 0.14 \\
			4245522468554091904 & BaII   & 1 & -2.67 & 0.12 & -0.19 & 0.15 \\
            \hline
		\end{tabular}
        \tablefoot{The solar abundances are from \citet{Asplund2009}. The full abundance table is available online at the CDS. }
  \label{tab:abundances}
\end{table*}

\section{Chemical abundance analysis}\label{sec:abundances}
\subsection{Procedure to measure abundances}
We derived the chemical abundance of stars using the 1D local thermodynamic equilibrium (LTE) spectral synthesis code MOOG \citep{Sneden1973a}. 
For each star, a model atmosphere was obtained by interpolating the grid of MARCS model atmospheres \citep{Gustafsson2008a} to the stellar parameters of the star. 
For most stars, we used photometric effective temperatures ($\teff$) and surface gravities ($\logg$); $\teff$ was obtained from $G-K_s$ colour, where $G$ and $K_s$ are the \textit{Gaia} DR3 and 2MASS magnitudes, respectively, using the relation from \citet{Mucciarelli2021a}.

The $\logg$ was obtained from the $K_s$ magnitude and the bolometric correction of \citet{Casagrande2014a} with an assumption of mass of $0.8\,M_\odot$.
In both cases, the photometries were dereddened using the extinction map of \citet{L22}.
The microturbulent velocity (\vt) was estimated by minimising the trend between the abundances derived from individual neutral iron lines and their strengths. 
We adopted $90\,\mathrm{K}$, $0.05$, and $0.12\,\kms$ as uncertainties for \teff, \logg, and \vt, respectively.\footnote{The uncertainty in \teff\, is mostly from uncertainties in the \teff-colour relation and uncertainties in extinction, the \logg\, uncertainty is equally affected by uncertainties in
the mass estimate, \teff, extinction and distance. For \vt\, uncertainty is determined by the range
of \vt\, which gives no significant trend between abundances and
strengths} In Table \ref{tab:stellar_params} we provide the values inferred for stellar parameters.
   
We find that the three stars show a significant ionisation imbalance in iron with the photometrically inferred stellar parameters. 
We thus determined the stellar parameters of these stars spectroscopically using the Python package \texttt{q2} \citep{Ramirez2014a}.
To keep the stellar parameters consistent with those of the other stars, we take a differential approach adopting reference stars. 
The three stars (and their corresponding reference stars) are ED-5-6039 (ED-3-2551), ED-3-5382 (ED-3-3089), and ED-3-4425 (ED-4-5195; see Table \ref{tab:obs} for the full \textit{Gaia} IDs). 

Reference stars were selected so that differences in effective temperatures and metallicities between the three target stars and their reference stars are small. We fix the stellar parameters of the reference stars and derive the relative line-by-line iron abundances for the three target stars. These are denoted $\Delta A_i({\rm Fe,I})$ for neutral lines and $\Delta A_i({\rm Fe,II})$ for ionised lines, where $i$ is the line index.
At each iteration, we adjusted the target star’s parameters as follows: the trend between excitation potential and $\Delta A_i({\rm Fe,I})$ is used to constrain $T_{\rm eff}$; the trend between reduced equivalent width and $\Delta A_i({\rm Fe,I})$ is minimised to determine $v_t$; and the difference between the median $\Delta A_i({\rm Fe,I})$ and $\Delta A_i({\rm Fe,II})$ is minimised to constrain $\log g$. The final adopted stellar parameters for each target are those that minimise these trends and differences.

We measured abundances based on equivalent widths of lines for Fe, Ca, Sc, Ti, Cr, Co, Ni, and Zn, and spectral synthesis for Li, C (CH), N (NH), Na, Mg, Al, Si, Mn, Cu, Sr, Y, Zr, Ba, La, and Eu.
We take hyperfine structure splitting into account for Li, Na, Al, Mn, Ba, and Eu.
In spectral synthesis, we consider the lines listed in Table \ref{table:linelist} and blending lines from the VALD database \citep{Piskunov1995a,Ryabchikova2015a}. 
After obtaining the best-fit spectrum, we synthesise an additional spectrum only with the lines of interest and measure equivalent widths through a direct integration for atomic features.
We applied line-by-line non-LTE corrections to the Na, Mg, and Al abundances by interpolating the grid of \citet{Lind2022}.

To obtain the final abundance of each element, we averaged the line-by-line abundances and estimated the uncertainties from the sample standard deviation of the line-by-line abundances ($\sigma$) and the number of lines ($N$) as $\sigma/\sqrt{N}$ when $N>3$; otherwise, we replace the $\sigma$ with that of neutral iron. 
We additionally consider the uncertainties due to the stellar parameters and add these in quadrature. 
The final abundances and their uncertainties are reported in Table~\ref{tab:abundances}.
When considering [X/Fe], we used the Fe abundance from the same ionisation stage as the species X.
For the species to which spectral synthesis is applied, the fitting result was visually inspected for each line to discard non-detections, poor solutions, and/or lines heavily contaminated by interstellar or telluric absorptions.

We also measure the radial velocities from the new high-resolution spectra and compare these with the \textit{Gaia} DR3 RVs, and similarly for the spectra taken from the archive. 
We check if any stars RVs vary by $> 5$ \kms. There are three stars that fall in this category: ED-2-4318 (BH3), ED-3-6557, and GE-2601. For ED-3-6557 and GE-2601 this RV variation is $\leq$ 7 \kms and less than 3$\epsilon\,(V_{\rm los})$ where $\epsilon\,(V_{\rm los})$ is the uncertainty on the \textit{Gaia} RV. However, the \textit{Gaia} RV uncertainty is sometimes inflated due to the presence of companions so could be hinting at a companion. 
ED-2-4318 (BH3 companion) also has a varying RV (50 \kms difference comparing \textit{Gaia} DR3 RV and the UVES RV quoted in the discovery paper) as already shown in \citet{Panuzzo:2024}. 

\subsection{Literature data comparisons}
To place the results in a wider context we compare our abundances to other literature datasets. This comparison is qualitative since small offsets between the datasets can be expected due to differences in the quality of spectra, instrumental set-up, choice of spectral synthesis code, line lists and derived stellar parameters.
We include the abundances from \citet{nissen2010,nissen2011} split into their high-$\alpha$ and low-$\alpha$ subsamples and \citet{reggiani2017} in our presentation of the abundances for elements that are in common between all studies. We also include abundances of the Helmi Streams and Sequoia from \citet{matsuno2022,matsuno2022high} that are homogenised to the same scale. This allows for a qualitative comparison with these low-$\alpha$ (slower enrichment) dwarfs. 

We also compared our abundances to those derived by \citet{Ceccarelli2024} for retrograde halo substructure, including ED-2, ED-3, and GE stars. We have 14 stars in common for the 12 elements presented in \citet{Ceccarelli2024} except for Y and Zn, where we have 11 and 10 stars respectively. Our comparison showed that the abundances are in good agreement, with a mean absolute difference of $\leq 0.1$ dex in the majority of elements. However, an exception is our \xfe{Al} abundances, which have an offset of 0.26 dex more negative. We discuss this further in Section \ref{sec:accreted_or_insitu}.

\section{Results}\label{sec:results}

\subsection{Metallicity distributions}\label{sec:metallicities}
We measure metallicity from the \ion{Fe}{II} lines, which are more reliable due
to their small sensitivity to the adopted stellar parameters and the non-LTE
effects. The derived homogeneous and precise metallicities of stars in each of the ED groups allowed us to obtain a detection of a metallicity spread, if present. This can be used to distinguish whether the progenitor is a dwarf galaxy \citep[with a metallicity dispersion $\geq 0.3\,\mathrm{dex}$, e.g.][]{Simon2019} or a star cluster \citep[negligible dispersion, dominated by measurement uncertainty $< 0.1\,\mathrm{dex}$, e.g.][]{Gratton2004}.

\begin{figure}
\centering
\includegraphics[width=0.39\textwidth]{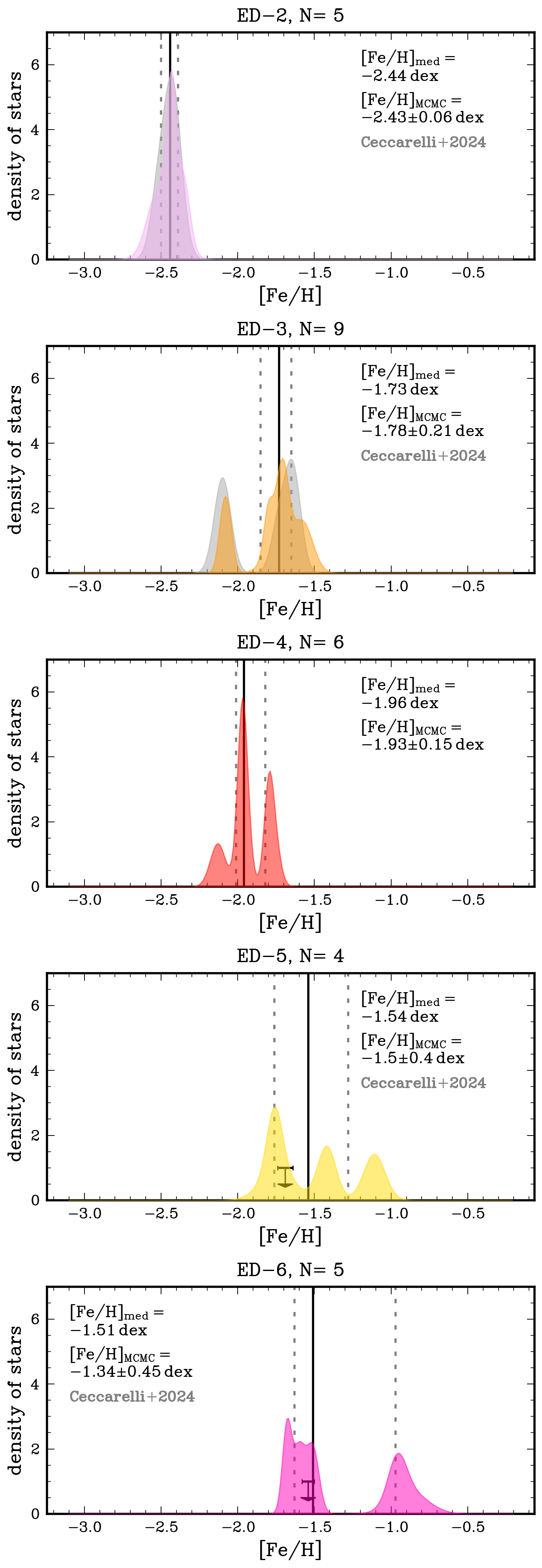}
\caption{Metallicity distributions of the different substructures coming from FeII lines. The median metallicity of the distributions are shown for each substructure, followed by the mean \xh{Fe} and intrinsic dispersion calculated with an MCMC assuming the metallicities follow a Gaussian distribution with a spread consisting of the individual metallicity uncertainties and an intrinsic dispersion. We also compared with \citet{Ceccarelli2024} metallicities where available. The comparison is shown as a distribution in grey or an arrow with an error bar where there is only one star. The distributions have been normalised for each sample such that integrating over metallicity will sum to 1. } 
\label{fig:MDFs}
\end{figure}

Figure~\ref{fig:MDFs} shows the metallicity distributions of the five substructures. The median [Fe/H] of each ED are quoted here. We could already see from the distributions that ED-2 and possibly ED-4 present a small dispersion in metallicity. ED-3 presents two groups in [Fe/H] at $\sim -2.1$~dex and $\sim- 1.7$~dex, as also seen in \citet{Ceccarelli2024}. From inspecting the orbital properties, there are no obvious dynamical differences in ED-3 stars that correlate with these two groups in [Fe/H], although we note that the \xh{Fe} $\sim- 1.7$~dex group contains seven stars, with the other only containing two stars. Both ED-5 and ED-6 present a larger spread in [Fe/H] and also the most metal-rich stars in the sample. Out of the four member stars of ED-5 two have [Fe/H] $\sim -1.7$~dex. One of which is ED-5-6039 is offset in $v_z$ by 100 \kms from the stream and is off the isochrone in the red sequence at $M_G \sim 0.5$ in Fig.~\ref{fig:CMDs}; the position in the CMD can be explained by our abundances and that this is a CN-rich star (see Sect. \ref{sec:results_Na}). The other ED-5 stars have [Fe/H] $\sim$ $-1.4$ and [Fe/H] $\sim$ $-1.1$~dex. The latter is ED-5-2653 which also has a high $\alpha$ and is the only star with positive $v_z$ that we have chemistry for, which is discussed further in Sect. \ref{sec:accreted_or_insitu}). ED-6 contains two groups, two stars at [Fe/H] $\sim$ $-1.0$~dex corresponding to those on the red sequence in Fig.~\ref{fig:CMDs} and including one with a high $\alpha$ (ED-6-3266) that is discussed further in Sect. \ref{sec:accreted_or_insitu}. The other group has a mean [Fe/H] $\sim -1.6$~dex containing three stars, one of which, ED-6-5617 has the opposite sign in $v_R$ to the majority of stream stars. Removing the high-$\alpha$ star from ED-6 we are left with three very tight in [Fe/H] at $\sim -1.6$~dex and one high metallicity star ED-6-6643 with low $\alpha$; however, we saw no obvious differences to exclude the high metallicity star from ED-6. 

\begin{figure*}
\centering
\includegraphics[width=0.9\textwidth]{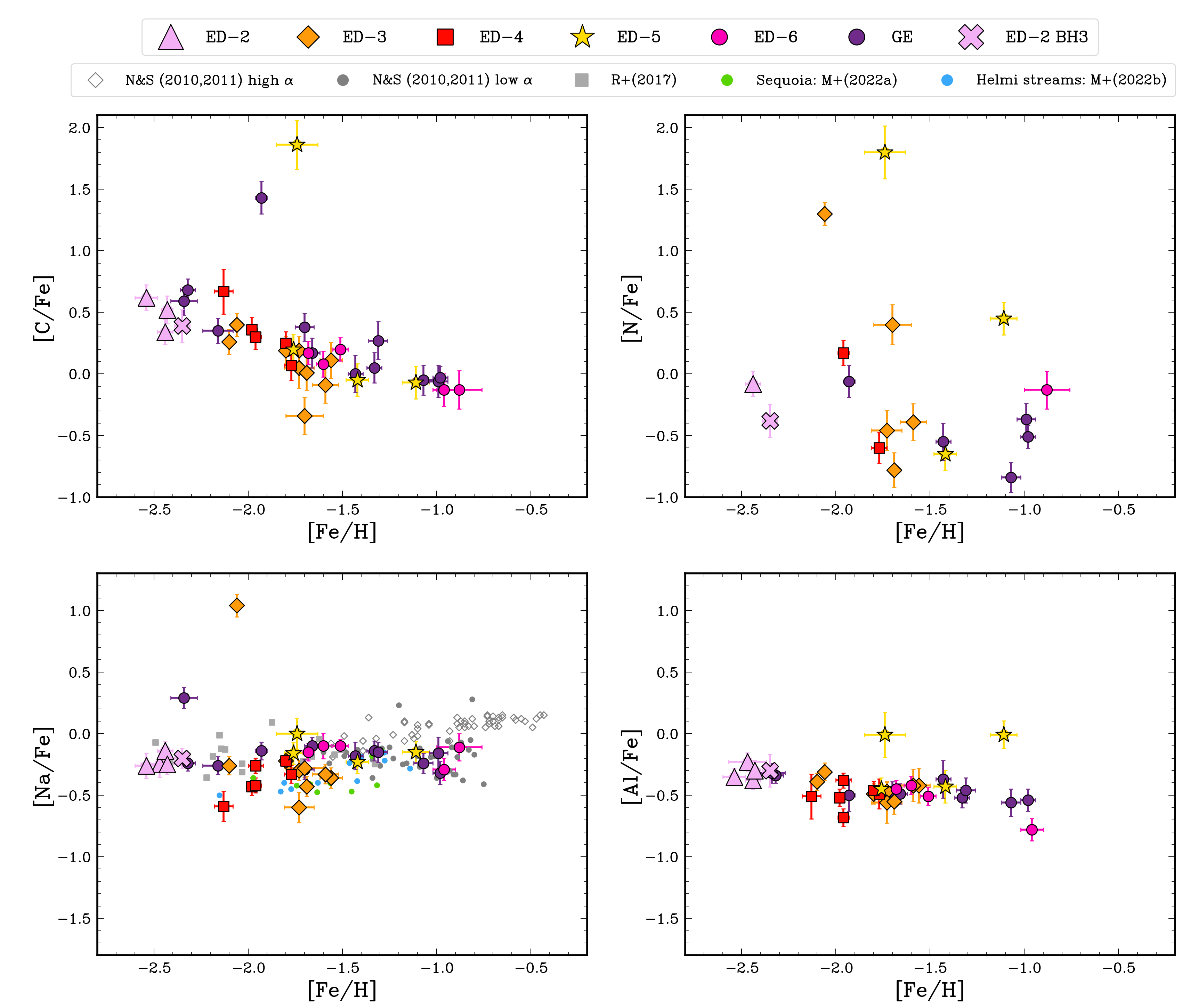}
\caption{Abundances of C, N, Al, and Na of the stars in the different ED groups and GE stars analysed homogeneously for comparison. The literature samples of halo stars from \citet{nissen2010,nissen2011}, \citet{reggiani2017}, for Sequoia from \citet{matsuno2022}, and the Helmi Streams from \citet{matsuno2022high} are shown for comparison where available. We note that N and Al are only measured for a subset of stars due to blended lines. } 
\label{fig:CN_Al_Na}
\end{figure*}

\begin{figure*}
\centering
\includegraphics[width=0.9\textwidth]{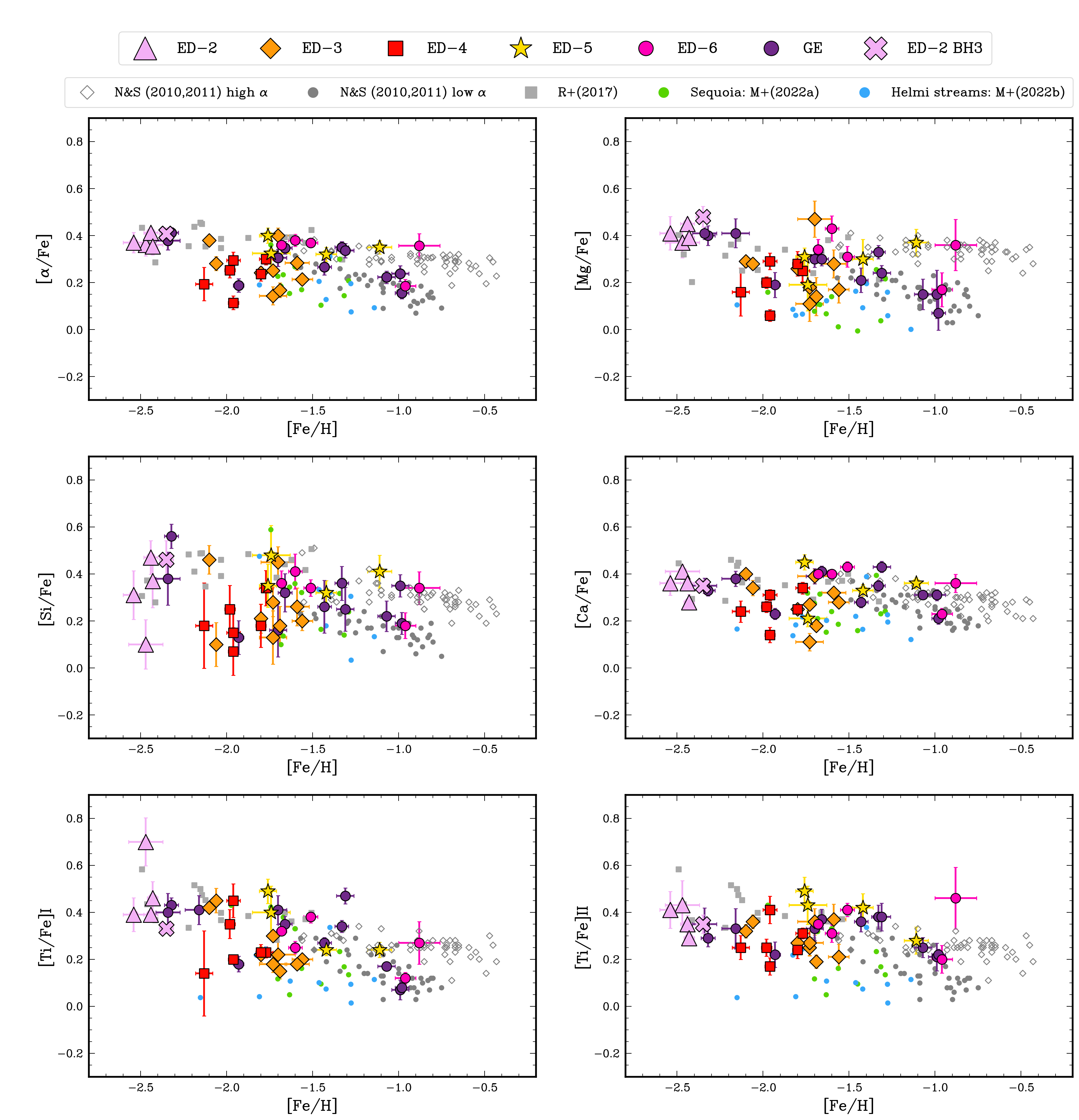}
\caption{Abundances of $\alpha$ of the different groups, including GE comparison stars (analysed homogeneously) in purple and a compilation of literature data of halo stars (as in Fig.~\ref{fig:CN_Al_Na}). The first panel shows the average $\alpha$ abundance, which is defined as $(\xfe{Mg}+\xfe{Si}+\xfe{Ca}+(\xfe{\ion{Ti}{I}}+\xfe{\ion{Ti}{II}})/2)/4$, an average over all the other panels. }
\label{fig:alpha}
\end{figure*}

To quantify the metallicities further we employ a Bayesian approach to measure the mean metallicity and the intrinsic metallicity dispersion ($\sigma_{\textrm{[Fe/H]},
int}$), assuming the stars follow a Gaussian distribution with a total $\sigma^2$ = $\sigma_{\textrm{[Fe/H]}, int}^2$ +
$\sigma^2_{\textrm{[Fe/H]}, i}$ , i.e. the sum in quadrature of an intrinsic
dispersion and the individual metallicity uncertainties. We maximised the log-likelihood, which has the following form
\begin{equation}
\begin{split}
\textrm{log}\,\, \mathcal{L}(\mu, \sigma_{\textrm{[Fe/H]}, int}) = -\frac{1}{2} \sum_{i} \Bigg[ \frac{(\text{[Fe/H]}_i - \mu)^2}{\sigma_{\textrm{[Fe/H]}, int}^2 +
\sigma_{\text{[Fe/H]},i}^2} + 
\\
\ln \left( 2 \pi [\sigma_{\textrm{[Fe/H]}, int}^2 + \sigma_{\text{[Fe/H]},i}^2] \right) \Bigg]
\end{split}
\label{eq:MCMC},
\end{equation}
and we performed Markov chain Monte Carlo (MCMC) sampling using \textsc{emcee} \citep{emcee} and a flat prior.
We found a best-fit metallicity and intrinsic dispersion for each group using the mean of the posterior distributions.  This more statistically robust measurement of the mean metallicity and dispersion (Eq. \ref{eq:MCMC}) is also presented in Fig.\ref{fig:MDFs} as [Fe/H]$_\textrm{MCMC}$. The figures in Appendix \ref{sec:appendix_corner_plots} show the corner plots of these MCMCs. ED-2 has a mean [Fe/H] of $-2.43$~dex with an upper limit on the intrinsic dispersion $\sigma_{\textrm{[Fe/H]}, int}$ of $0.13$~dex. ED-3 has a mean [Fe/H] of  $-1.78$~dex with a resolved dispersion of $\sigma_{\textrm{[Fe/H]}, int}$ of $0.21$~dex, driven by the two metal-poor stars at [Fe/H] $\sim$ -2.1 dex. 
ED-4 has a mean [Fe/H] of $-1.93$~dex with a dispersion of $\sigma_{\textrm{[Fe/H]}, int}$ of $0.15$~dex. ED-5 and -6 have means of $-1.51$ and $-1.35$~dex respectively, with the most significant dispersions of 0.41 and 0.45~dex respectively. This is comparable to the GE stars which have a mean [Fe/H] of $-1.6$~dex and a dispersion of 0.53~dex. We interpret these results in a broader context in Sect.~\ref{sec:GC_or_dwarf}.

\subsection{C, N, and odd-$Z$ elements}\label{sec:results_Na}
\begin{figure*}
\centering
\includegraphics[width=0.85\textwidth]{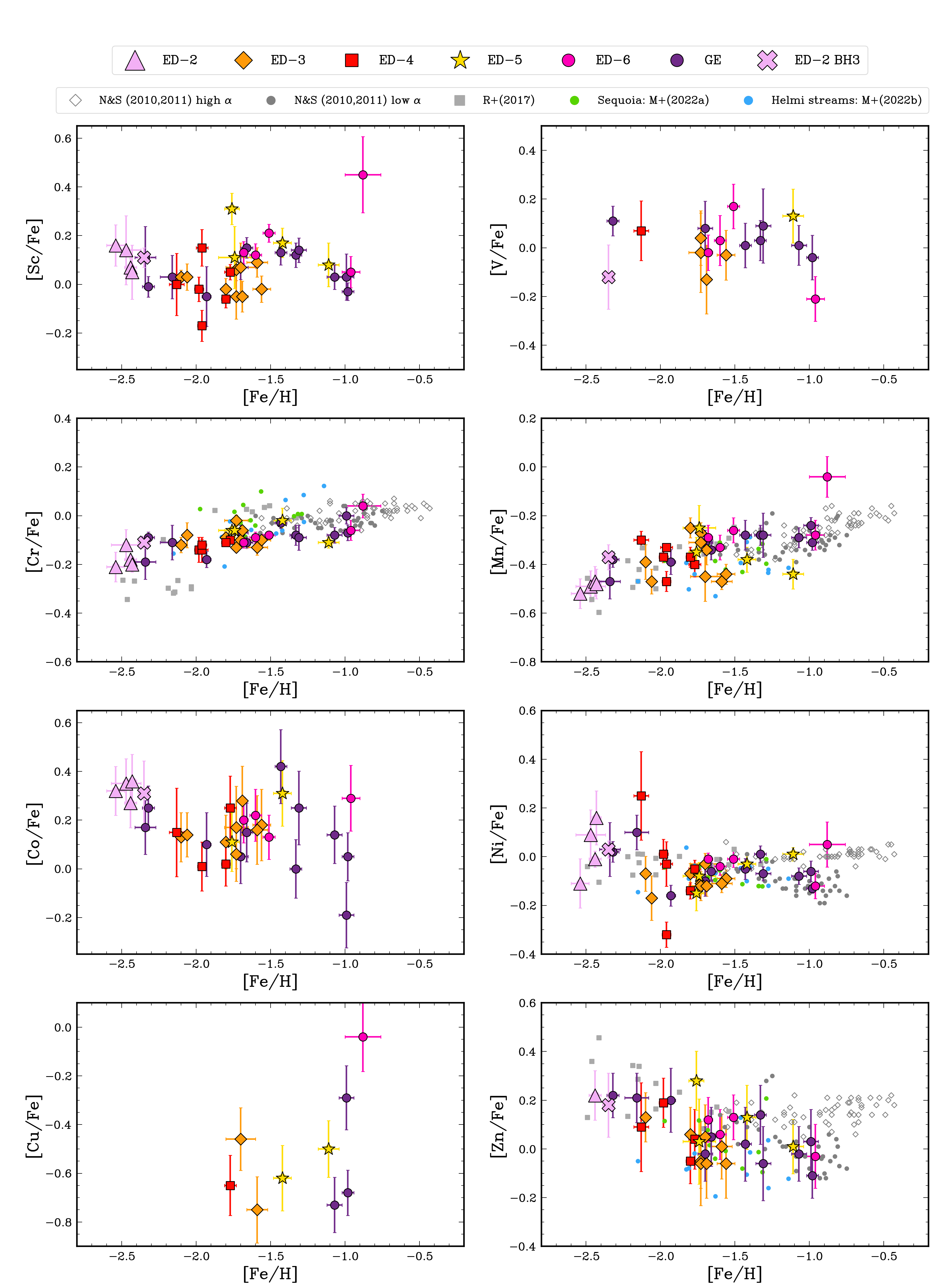}
\caption{Same as shown in Fig.~\ref{fig:CN_Al_Na} but for Iron peak elements.} 
\label{fig:Fe_peak}
\end{figure*}
Figure \ref{fig:CN_Al_Na} presents abundances of C, N, Na, and Al of stars as a function of \xh{Fe}.
While no stream stands out in any of the abundance ratios, there are a few outliers in each elemental abundance. 
ED-3-5382 has a low \xfe{C} compared to stars with similar metallicity. 
Since this star is the most evolved star in this sample with $\logg=1.51$, the low \xfe{C} is likely due to internal nucleosynthesis and mixing within the star. 
Indeed the evolutionary correction by \citet{Placco2014} predicts that its \xfe{C} has been reduced by 0.42~dex from the main-sequence value.
This correction would bring up the \xfe{C} of ED-3-5382 to the value comparable to other stars. 
Contrary to ED-3-5382, both GE-2601 and ED-5-6039 have high C abundance with $\xfe{C}>1$. 
They are also enhanced in $s$-process element abundances with $\xfe{Ba}>1$, and thus we consider them carbon-enhanced metal-poor stars with $s$-process enhancements \citep[CEMP-s stars][]{Beers2005}. The high C abundance could be related to why ED-5-6039 is off in the CMD (Fig.~\ref{fig:CMDs}).
Such stars are usually considered to have accreted material rich in C and $s$-process elements from more evolved companions that have undergone the asymptotic giant branch phase. As already mentioned GE-2601 shows an RV variation of $\sim$ 7 \kms and the error on the \textit{Gaia} DR3 RV is 6 \kms. It does not have an earlier RV measurement in \textit{Gaia} since it has a \verb|phot_g_mean_mag| fainter than 13. While the star has a \verb|RUWE| below 1.4 given our definition of our sample selection,
the large RV uncertainty in \textit{Gaia} DR3 suggests that GE-2601 is likely a binary. On the other hand, ED-5-6039 has no RV variation and a lower \textit{Gaia} DR3 RV uncertainty, it also has no earlier epoch RV from \textit{Gaia} to be able to investigate further. 
Future multi-epoch RV measurements are needed to reveal if there is a RV variation in ED-5-6039. Furthermore, if the binary is face-on we would not expect an RV variation so this alone does not rule out that this star could be a binary. However, we note that ED-5-6039 is likely not a true ED-5 member, given that is offset to more negative $v_z$ by 100 \kms from the negative $v_z$ stream. 

\begin{figure*}
\centering
\includegraphics[width=0.87\textwidth]{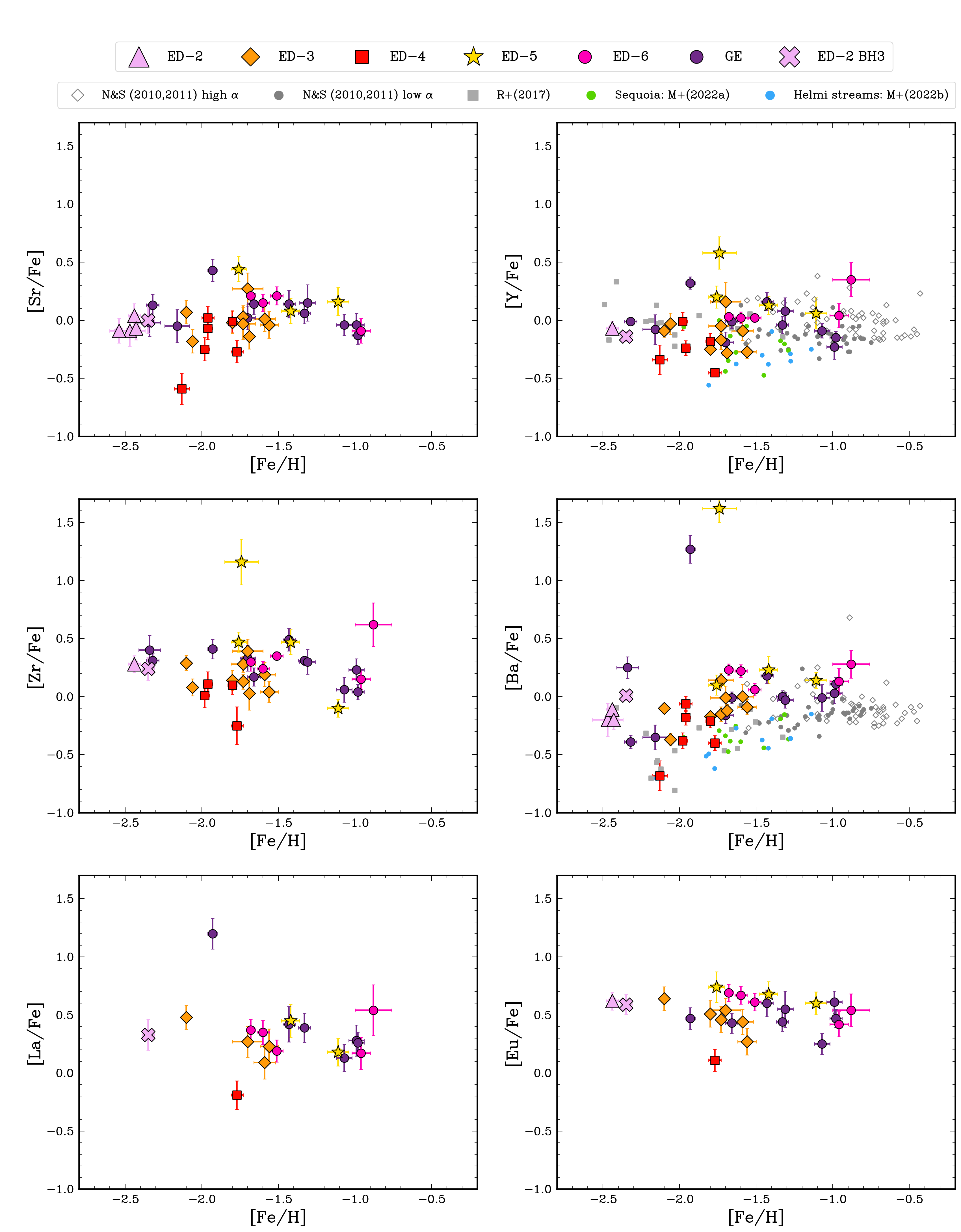}
\caption{Same as shown in Fig.~\ref{fig:CN_Al_Na} but for neutron capture elemental abundances.} 
\label{fig:heavy_elements}
\end{figure*}

Nitrogen abundances are derived from the NH feature at 336\,nm. 
Since the molecular feature is not very strong in warm stars and since the signal-to-noise ratio tends to be low at this wavelength, nitrogen abundances are available for only a small subset of stars.
We thus focused on stars with extremely high \xfe{N} to avoid the bias we have against those with low \xfe{N}.
There are two stars with \xfe{N}$>1$ (ED-3-6334 and ED-5-6039). 
As we discussed above, ED-5-6039 also exhibits high C and although there is no evidence of a binary from the radial velocity we cannot rule out a face-on binary, which could explain the high C and N abundance of ED-5-6039. For ED-3-6334 there is also no indication of the presence of a binary companion to this star, we find a RV variation between \textit{Gaia} DR3 and that measured from our spectra of $\sim$ 3.6\, \kms with a \textit{Gaia} DR3 RV error of 3\, \kms, the star has a \texttt{RUWE} of 1.13 and since it is fainter than \verb|phot_g_mean_mag| of 13 there is no RV information from earlier \textit{Gaia }data releases. The same caveat applies that a face-on binary cannot be ruled out with the RV variation alone.
We also note that this accompanies a large Li enhancement, which is challenging to explain without assuming past interaction with a companion/companions \citep{matsuno2025}.
We also note that the apparent large dispersion in \xfe{N} could be due to stellar evolution as a conversion of a small amount of C to N can lead to large N enhancements without observation of significant C depletion. 

While C and N can be strongly affected by stellar evolution,  Na and Al are less likely to be.
The two stars with high Na abundance, $\xfe{Na}>0$, are GE-6155 and ED-3-6334. 
While the origin of these Na enhancements is unclear, that of ED-3-6334 could be related to its Li and N enhancements, as discussed above. 
There seems to be variation in the average \xfe{Na} ratios of the ED streams, and some ED streams seem to have different \xfe{Na} ratios than the GE stars at similar metallicity: while ED-3 (after excluding ED-3-6334) and some ED-4 stars have lower \xfe{Na} ratios than GE stars at similar metallicities, ED-2, ED-5, and ED-6 seem to follow the GE sequence. 
We interpret this feature in Section \ref{sec:accreted_or_insitu}. 

Compared to Na abundances, Al shows a tighter sequence and smaller variation.
Although there is an outlier (ED-5-6039) and a scatter at $\xh{Fe}>-1.2$, this likely reflects the difficulty in the analysis. 
The Al abundance measurements are based on \ion{Al}{I} lines at 394 and 396 nm.
These lines are in the vicinity of Ca HK lines, can be blended with molecular lines, and become very strong at high metallicity. 
Even though we take blends into account, these difficulties prevent us from precisely measuring Al abundances in stars with high C abundance (this is the case for ED-5-6039) or those with high metallicity. 
Despite these limitations, we do not observe obvious differences in \xfe{Al} among ED streams or between ED streams and GE stars.

\subsection{$\alpha$-elements}\label{sec:alpha}

\fig{fig:alpha} presents trends of $\alpha$-elements. 
The abundance ratio, \xfe{\alpha}, can be used as an indicator of the timescale of chemical enrichment and star formation efficiency.
While massive stars disperse both $\alpha$ elements and Fe in a short timescale, Fe is also produced with a significant amount by type~Ia supernovae.
Since type~Ia supernovae are explosions of white dwarfs, the end products of low-mass star evolution, as a result of binary interaction, they start happening long after the onset of star formation.
Thus, a system with a long star formation timescale tends to show low \xfe{\alpha}.
Unlike lighter elements, there are no obvious outliers in the abundance trends of $\alpha$-elements, and each stream shows a sequence or clump, except ED-6. 
There are variations in \xfe{\alpha} among different streams. 
ED-3 and ED-4 tend to show lower $\alpha$-element abundances compared to GE stars or other ED streams at similar metallicities and overlap more with the region occupied by Sequoia and the Helmi Streams \citep{matsuno2022,matsuno2022high}. ED-6-3266 and ED-5-2653 are more consistent with the high-$\alpha$ track (see high-$\alpha$ stars from \citealt{nissen2010,nissen2011}), having higher \xfe{\alpha} at fixed \xh{Fe} than GE stars indicative of an in situ origin (e.g. see the \xfe{Si} and \xfe{Mg} abundances in Fig.~\ref{fig:alpha}). ED-5-2653 is the only star for which we have chemical information in the positive $v_z$ stream of ED-5, which is only made up of fewer than one-third of the ED-5 stars.
One GE star, GE-1159, could be consistent with the high-$\alpha$ track in \xfe{Si} but more consistent with low $\alpha$ in \xfe{Mg} and the average $\alpha$.
We note that ED-2 stars present mostly a clump in $\alpha$ abundances with little spread and the BH3 companion being consistent with all. One star, ED-2-4245, is slightly lower in \xfe{Si} and higher in \ion{\xfe{Ti}}I but consistent within all other \xfe{\alpha}.
We interpret these abundance differences more throughout Sect. \ref{sec:discussion}.

\subsection{Iron-peak elements}\label{sec:fe_peak}

\fig{fig:Fe_peak} shows abundance trends for iron-peak elements. 
Some of the elements, such as Sc, Ni, and Zn, have large contributions from massive stars but little contributions from type~Ia supernovae and behave similar to $\alpha$-elements.
Indeed, Sc, Ni, and Zn abundances of ED-3 provide a consistent picture with the $\alpha$-elements, in the sense that its stars show lower \xfe{X} in these elements. 
It is not as clear for ED-4, but this more likely reflects the difficulty of measuring the abundances of these elements at low metallicity as seen from the large error bars. 
All of the streams and the comparison samples appear to follow similar trends in other iron-peak elements, V, Cr, Mn, Co, and Cu, within the current measurement uncertainty. 

\subsection{Neutron-capture elements}\label{sec:neutron_capture}

\fig{fig:heavy_elements} shows abundances of neutron-capture elements. 
The two aforementioned suspected CEMP-s stars (GE-2601 and ED-5-6039) appear as outliers in almost all the panels reflecting nucleosynthesis in their companions. These stars have very high [Ba/Y] with GE-2601 also presenting high [Ba/Eu], typical of binary mass transfer, while ED-5-6039 does not have a Eu measurement due to the high C. 
We remove these two stars in the interpretation of neutron-capture elements abundances. 
Other than the two outliers, most ED streams show a tight scatter in the abundances of neutron-capture elements.

All stars in the ED streams have higher [Ba/Fe] than the literature compilation \citep{nissen2010,nissen2011,reggiani2017,matsuno2022,matsuno2022high}, which is indicative of an offset. This is present in both  [Ba/Fe] and [Ba/H]. We note that there is a difference in the Ba linelist between that adopted by \citet{nissen2010,nissen2011} and our analysis of the ED streams, and that all literature abundances we are using were homogenised to be on the same scale as  \citet{nissen2010,nissen2011}. We adopted $\sim 0.1$ dex smaller gf values,\footnote{The gf value is related to the probability of an electron to transition between the given energy levels and therefore directly related to the intensity of that line in a spectrum} which would result in a 0.1 higher [Ba/H] abundance. As only a few stars overlap between the samples, we did not quantify this offset but caution that it is likely a result of systematics and not real. Homogenising our abundances with the literature compilation would be an obvious next step for future work.

We note that there is also variation among different streams: while ED-5 and ED-6 tend to show high neutron-capture element abundances, ED-3 and ED-4 tend to show lower abundances. In general ED-3 and ED-4 appear to have lower \xfe{Sr} and \xfe{Y} at fixed \xh{Fe} than GE, as was seen previously in the Helmi streams and Sequoia (\citealt{aguado2021,matsuno2022,matsuno2022high} and also in Fig.~\ref{fig:heavy_elements}). 
Particularly ED-4-5195 stands out with $\xfe{Sr}<-0.5$ and $\xfe{Ba}<-0.5$ with a mild deficiency in $\xfe{Y}$ of -0.34. 
Zr, La, and Eu lines were not detected for this star. 
While ED-4-2436 also shows low abundances in almost all the elements, the low abundance is particularly clear in \xfe{Y}, \xfe{Zr}, \xfe{La}, and \xfe{Eu}.

\section{Discussion}\label{sec:discussion}

Throughout Section \ref{sec:results} we have presented chemical abundances measured for the ED streams and highlighted the main trends when comparing with each other and with homogeneous GE abundances. We have presented in Sect. \ref{sec:metallicities} the metallicity distributions and shown that for ED-2 we do not detect an intrinsic dispersion in \xh{Fe}. For all other EDs we do detect a dispersion with ED-5 and ED-6 presenting the largest dispersion and highest mean \xh{Fe}. ED-5 and ED-6 appear to follow the GE stars in all chemical abundances, except for a couple of outlier stars. 
Both ED-3 and ED-4 present low [$\alpha$/Fe] abundances, similar to the trends seen for smaller dwarfs, the Helmi streams and Sequoia \citep{matsuno2022,matsuno2022high}. There are also hints that ED-3 and ED-4 present lower [Sr/Fe] and [Y/Fe] abundances at fixed \xh{Fe}, which is also consistent with what has been seen for the Helmi streams and Sequoia. Here we discuss these findings in more detail as well as the implications for the nature of the progenitors of these streams.

We presented the \xfe{\alpha} abundances of the streams in  Fig.~\ref{fig:alpha}, showing that the majority of the streams contain stars consistent with the low \xfe{\alpha} regime, where stars of accreted origin are expected. ED-5 and ED-6 have higher metallicity stars that populate the region where the accreted and in situ tracks separate more clearly. From this, we can say that both streams contain stars that are consistent with the low-$\alpha$ track and with the GE abundances. However, both ED-5 and ED-6 have one star (ED-5-2653 and ED-6-3266), out of the total 4 and 5 stars respectively, which are more consistent with the higher \xfe{\alpha} track. These two stars are at considerably higher \xfe{\alpha} at fixed \xh{Fe} than GE stars. The ED-5 star could be potentially ruled out as a stream member based on having an opposite $v_z$ to the other stars for which we have chemical abundances. Follow-up of more stars in the positive $v_z$ stream of ED-5 is needed to verify this. The other ED streams lie at lower metallicity where the separation of in situ and accreted stars with \xfe{\alpha} is less clear. However, we could see that both ED-3 and ED-4 present substantially low-$\alpha$ abundances, lower than GE at fixed \xh{Fe}, which points to an accretion origin but also suggests that they originated in a smaller dwarf, similarly to the chemical patterns seen for the Helmi streams and Sequoia \citep{matsuno2022,matsuno2022high}. ED-2 is the lowest metallicity stream in the sample and is relatively compact in \xfe{\alpha}. The position in \xh{Fe}-\xfe{\alpha} space alone does not allow us to discern between an accreted or in situ origin. 

The first panel of Fig.~\ref{fig:accreted} shows the [Al/Fe]--[Mg/Mn] plane, which is commonly used to distinguish accreted (chemically unevolved) from in situ (chemically evolved) populations \citep[as proposed by][see also \citealt{fernandes2023comparative}]{das2020ages}. This separation (based on APOGEE data) is indicated by the black dashed line with accreted or unevolved stars exhibiting lower [Al/Fe] at similar [Mg/Mn] to in situ stars. However, caution must be taken because this separation has been shown to be robust at intermediate \xh{Fe} (i.e. GE) but is less well studied for lower metallicities, and it is suggested that both in situ and accreted metal-poor stars will overlap in the high [Mg/Mn], low [Al/Fe] region \citep[hence the term unevolved; see][]{fernandes2023comparative}. For comparison, we show here a sample of APOGEE DR17 data in the background, selecting high-quality stars with \verb|FE_H_ERR| $< 0.1$, \verb|ALPHA_M_ERR| $< 0.2$, \verb|S/N| $> 70$, \verb|EXTRATARG| $=0$, \verb|STARFLAG| $=0$ and \verb|ASPCAPFLAG| $=0$. There appears to be an offset between our abundances and APOGEE DR17, in the sense that we measure lower [Al/Fe] abundances. We only have two stars in common with APOGEE DR17 that have \xfe{Al} abundances, one is $\sim$0.2~dex higher in \xfe{Al} and the other $\sim$0.1~dex lower when compared with APOGEE, but due to the limited sample it is difficult to draw a conclusion from this. We do, however, have a larger overlap with the stars in \citet{Ceccarelli2024}. We have 14 stars in common for which we measure \xh{Fe} and find a negligible average offset of $-0.04$~dex when comparing to our \ion{[Fe/H]}{II} or $-0.12$~dex when comparing to our \ion{[Fe/H]}I. For the \xfe{Al} abundances we have 13 stars in common (across all metallicities) and find a more significant offset, our Al abundances are always lower with a mean offset in \xfe{Al} of $-0.26$~dex. Since the abundances presented in \citet{Ceccarelli2024} agree well with APOGEE DR17, across all metallicities, it suggests that we have an offset in \xfe{Al} with APOGEE also. Shifting our abundances by $\sim -0.3$~dex in \xfe{Al} would result in some stars lying near the accreted or in situ boundary seen in Fig.~\ref{fig:accreted}. However, even with these offsets, we can observed that all of the ED stars, except ED-5-6039, have chemical abundances that fall into the accreted or chemically unevolved region. ED-5-6039, that can be seen at [Al/Fe] $\sim$ 0 and [Mg/Mn] $\sim$ 0.4 in our scale, is the star with high \xfe{C} and high [Ba/Y] that could be a CEMP-s star. ED-5-2653 is the high-$\alpha$ star seen in ED-5 and is the other star with [Al/Fe] $\sim$ 0 but higher [Mg/Mn] $\sim$ 0.8. The ED-6 star that has a high $\alpha$ (ED-6-3266) falls within the accreted or unevolved region of this diagram.

\subsection{Accreted or in situ}\label{sec:accreted_or_insitu}
\vspace{-0.13cm}
\begin{figure}
\centering
\includegraphics[width=0.41\textwidth]{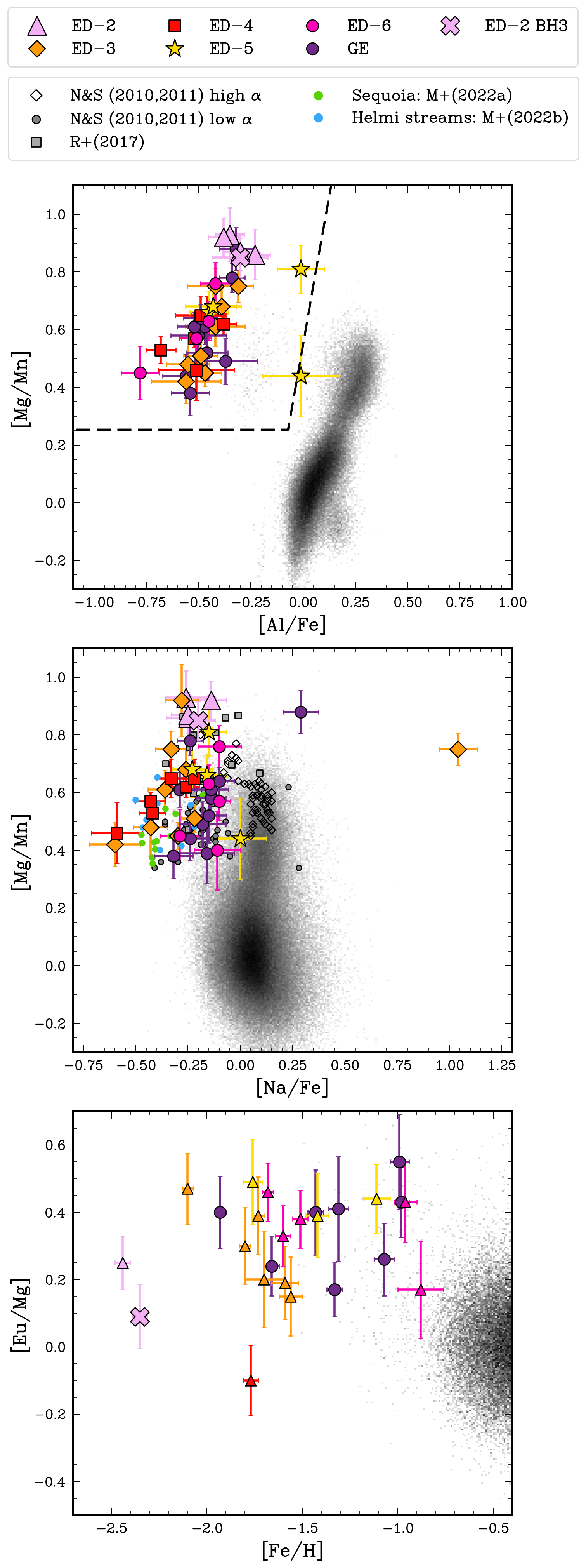}
\caption{Chemical abundance spaces commonly used to distinguish between accreted and in situ stars. The top panel shows [Mg/Mn] as a function of the \xfe{Al} abundances, with the APOGEE DR17 control sample in the background (see the text for details). The dashed line shows the region typically populated by accreted stars. The middle panel shows [Mg/Mn] as a function of the \xfe{Na} abundances. The bottom panel shows [Eu/Mg] as a function of \xh{Fe}. The bottom two panels show GALAH DR3 data in the background since GALAH provides \xfe{Na} and \xfe{Eu} measurements. } 
\label{fig:accreted}
\end{figure}

\begin{figure*}
\centering
\includegraphics[width=0.9\textwidth]{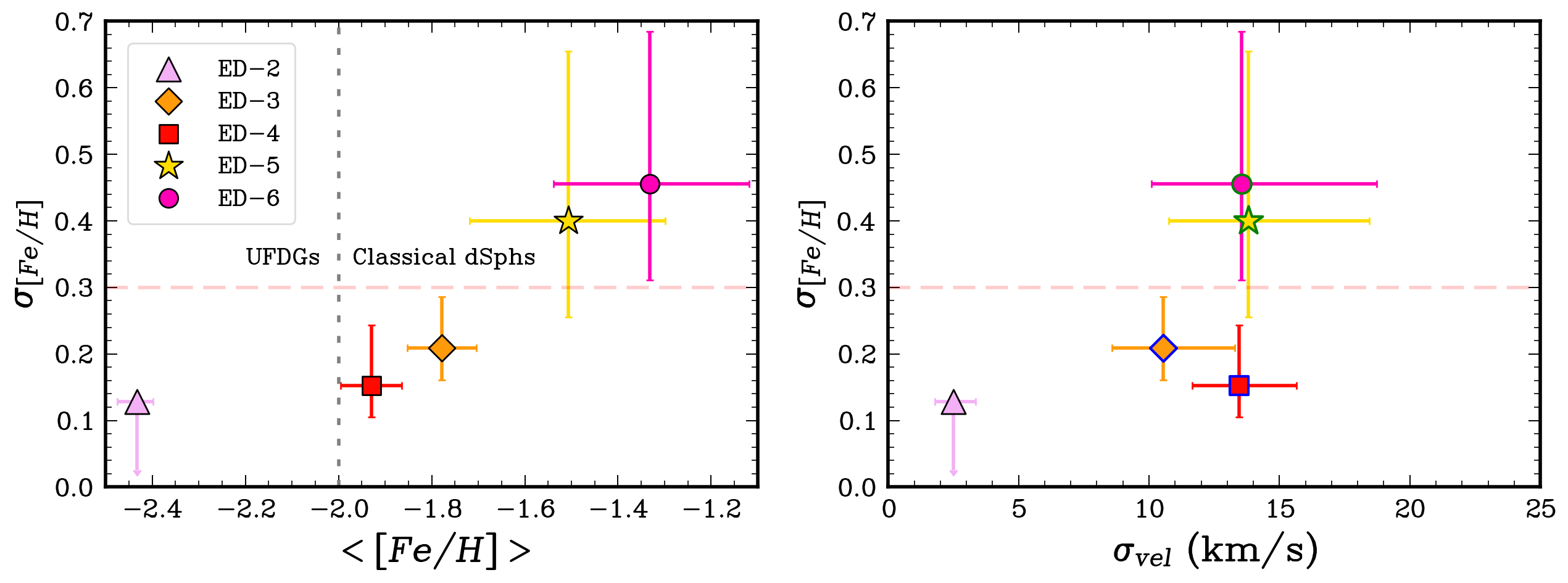}
\caption{Left panel: Mean metallicity and dispersion taken from the MCMC fit. Errors are taken from the 16th and 68th percentiles of the samples, where a dispersion is resolved; otherwise, an upper limit is shown at the 84th percentile, as seen for ED-2. The grey dashed line in the left panel corresponds to a [Fe/H] of $-2$ dex, the typical separation between classical dwarf spheroidals and ultra-faint dwarf galaxies. Right panel: FeII dispersion against the velocity dispersion. For the streams with multiple clumps (all except ED-2) in velocity space, the velocity dispersions were calculated from the clump with more than four members. For those outlined in green, the velocity clump from which the dispersion was calculated corresponds to the negative $v_z$ (or $v_x$ for ED-6) stream, and those outlined in blue correspond to the positive $v_z$ stream. The red dashed line in both panels at $0.3\,\mathrm{dex}$ represents the typical minimal \xh{Fe} spread seen in dwarf galaxies \citep{Simon2019}. } 
\label{fig:met_spread}
\end{figure*}

The second panel of Fig.~\ref{fig:accreted} presents the same space but instead of \xfe{Al}, we use here \xfe{Na}. Both are odd Z elements produced during SNe II and the production of these elements are expected to be similar, depending on the C+N abundance \citep{kobayashi2006galactic}. In GALAH Al is not well measured at low \xh{Fe} and so Na is preferred \citep{buder2022galah}. Therefore, for comparison, we show here a GALAH DR3 control sample selected using the following criteria; \verb|flag_sp| $=0$, \verb|flag_fe_h| $=0$, \verb|flag_Mg_fe| $=0$, \verb|flag_Mn_fe| $=0$, \verb|flag_Eu_fe| $=0$, \verb|flag_Na_fe| $=0$, \verb|snr_c3_iraf| $>30$, \verb|e_Na_fe| $\leq0.2$, \verb|e_Mg_fe| $\leq0.2$, \verb|e_Mn_fe| $\leq0.2$, \verb|e_Eu_fe| $\leq0.2$, and \verb|e_fe_h| $\leq0.2$. We show our abundances and also the abundances from literature halo samples that have \xfe{Na} too; namely from \citet{nissen2010,nissen2011}, \citet{reggiani2017}, \citet{matsuno2022}, \citet{matsuno2022high}. We observed that the GALAH control sample mostly populates the high \xfe{Na}, low [Mg/Mn] region which is dominated by highly evolved low-$\alpha$ thin disc stars. The high-$\alpha$ in situ halo stars populate the higher [Mg/Mn] region, as populated by the \xfe{Na} > 0 stars from \citet{nissen2010,nissen2011}. We note that \citet{buder2022galah} discuss in detail the limitations of selecting purely accreted stars in the [Na/Fe]--[Mg/Mn] space, and low metallicity in situ stars can present low \xfe{Na} and overlap with the accreted region. Some ED stars (primarily ED-5 and ED-6) follow the GE stars distribution at low \xfe{Na}. Sequoia and the Helmi streams present lower \xfe{Na} than GE, which is also seen for ED-4 and ED-3 stars. The two stars that are enhanced in \xfe{Na} are ED-3- 6344 and GE-6155, as presented already in Sect. \ref{sec:results_Na}. Again, ED-5-6039 lies close to the region that is populated by in situ stars.

The bottom panel of Fig.~\ref{fig:accreted} presents [Eu/Mg] as a function of \xh{Fe}. It has been suggested that the r-process element europium (Eu) could be a potential chemical tag to resolve differences between in situ and accreted populations \citep[e.g.][and more recently \citealt{monty2024ratio}]{recio2021heavy}  that the [Eu/$\alpha$] is different between accreted and in situ halo stars, across all \xh{Fe} \citep[see also][]{matsuno2021r}. The power of Eu can be explained by considering that GCs show light element abundance variations in the elements commonly used to distinguish between accreted and in situ populations. However, Eu appears to be unaffected by these chemical processes that are internal to GCs \citep[e.g.][]{roederer2011primordial,recio2021heavy,Kirby2023}. Since GALAH DR3 also provides \xfe{Eu} measurements we show here our [Eu/Mg] abundances for ED stars and GE stars compared with the GALAH control sample. Here the same quality cuts are applied as before but instead of \verb|e_Na_fe| <= 0.2, \verb|e_Mn_fe| <= 0.2, we use \verb|e_Eu_fe| <= 0.2. It is important to note that GALAH is unable to measure Eu at low \xh{Fe} and our own abundances are also not complete. We can only measure Eu at low \xh{Fe} if it is high, creating a bias in our presented abundances. All stars with \xh{Fe} $> -1.7$~dex have a Eu abundance (which means we measure Eu for all ED-5 and ED-6 stars, but we are missing stars from the other groups), for \xh{Fe} $< -1.7$ there is a bias to higher Eu.  
The bottom panel of Fig. \ref{fig:accreted} shows that the GALAH control sample is dominated by in situ stars and also high metallicity stars, at least in part because of the limitations in measuring Eu at low \xh{Fe}. We could see that ED-6-3266, the high-$\alpha$ star, presents a low [Eu/Mg] that could be more consistent with the in situ stars. 
It is unclear whether there are offsets in our data compared to those presented in \citet{monty2024ratio}, but the abundances suggest that most stars in the ED streams have high [Eu/Mg], which is consistent with the trends that have been seen for GE and proposed for accreted populations~in~general.

\subsection{Globular cluster versus dwarf galaxy origin}\label{sec:GC_or_dwarf}

In the first panel of Fig.~\ref{fig:met_spread} we show the mean metallicity and intrinsic metallicity dispersion obtained through MCMC fitting (described in Sect. \ref{sec:metallicities} for the ED groups). This can be used to distinguish whether the progenitor is a dwarf galaxy \citep[with a metallicity dispersion $\geq 0.3\,\mathrm{dex}$, e.g.][]{Simon2019} or a star cluster.
Globular clusters present a negligible spread in \xh{Fe}, typically smaller than $ 0.1\,\mathrm{dex}$ \citep[e.g.][]{Gratton2004}. This figure shows that ED-2 has a very small \xh{Fe} dispersion, as seen previously in \citet{balbinot2024} who demonstrates in detail why ED-2 is likely a disrupted star cluster. 
These authors were only able to place an upper limit on the spread of light elements in ED-2. 
We conclude the same from our abundances of 5 ED-2 stars. The Mg-Al abundances do appear to show hints of an anti-correlation but it is not significant when we consider the uncertainties. It is known that metal-poor GCs tend to have the smallest spread in light element abundances typical of these systems \citep[e.g.][]{Carretta:2009, Gratton2019}, and so we should not rule out that ED-2 could have been a GC since ED-2 lies around what is known as the metallicity floor for GCs (\xh{Fe} $\sim$ -2.4), below which no GCs have been observed. Larger samples of ED-2 stars are needed to shed more light on this. 

For ED-3 and ED-4 the mean metallicity is higher and a metallicity dispersion is detectable, although it is still below 0.3~dex, leaving the interpretation from \xh{Fe} alone ambiguous. Both ED-5 and ED-6 present a high mean metallicity and significant dispersion, even if we remove the high-$\alpha$ stars contaminating these samples. All the evidence for ED-5 and ED-6 suggests that these stars belong to a larger dwarf galaxy, given the scatter, high \xh{Fe} and the large spread in the CMD indicating a spread in populations. The abundances presented throughout this paper point to an association with GE (after removing the high-$\alpha$ stars), suggesting that ED-5 and ED-6 correspond to high-energy tails of its extended debris.

Since the nature of ED-3 and ED-4 remains ambiguous, we examine in more detail their dynamics. We can look at the velocity dispersion, given that streams of disrupted GCs have lower velocity dispersions than those of a dwarf galaxy \citep[e.g.][]{bonaca2024stellar,li2022s}. If a stream is sufficiently long then it can wrap around and pass through the solar neighbourhood multiple times, with stars on different phases of the orbit being present in the form of multiple clumps in velocity space, within the local volume. For several of the ED streams (all but ED-2) we observe multiple clumps in velocity space and refer to each of these velocity clumps as individual streams within the ED substructure. 
For the ED's which present multiple streams, one stream contains most of the stars and the other only $\sim$3-4 stars. To measure the velocity dispersion we, therefore, selected stars within the velocity clump that contains a sufficient number of stars. For ED-3 and Ed-4 this is the stars with $v_z >0$, for ED-5 the stars with $v_z<0$ and for ED-6 this is $v_x<0$. ED-2 presents only a single clump in velocity space.

The velocity dispersions are calculated by performing a principal component analysis (PCA) decomposition in 3D velocity space on the selected stars.
We took the PCA component with the lowest dispersion amplitude and model the projected velocities and their individual uncertainties to obtain an estimate for the intrinsic dispersion, this is done in the same way as the metallicities in Sect. \ref{sec:metallicities} equation \ref{eq:MCMC}. The velocity dispersion is plotted against the metallicity dispersion in the second panel of Fig.~\ref{fig:met_spread}. ED-2 has a low velocity dispersion, consistent with a colder stream. ED-3 and ED-4 do present higher velocity dispersions. However, the PCA decomposition does not take into account any velocity gradients that might be present and velocity dispersion increases as the orbit approaches a turning point \citep[e.g.][]{helmi1999a}. Since all of the ED streams are approaching pericentre $\sigma_{vel}$ is at best suggestive but not a fully robust indicator of progeny.

\begin{figure*}
\centering
\includegraphics[width=0.98\textwidth]{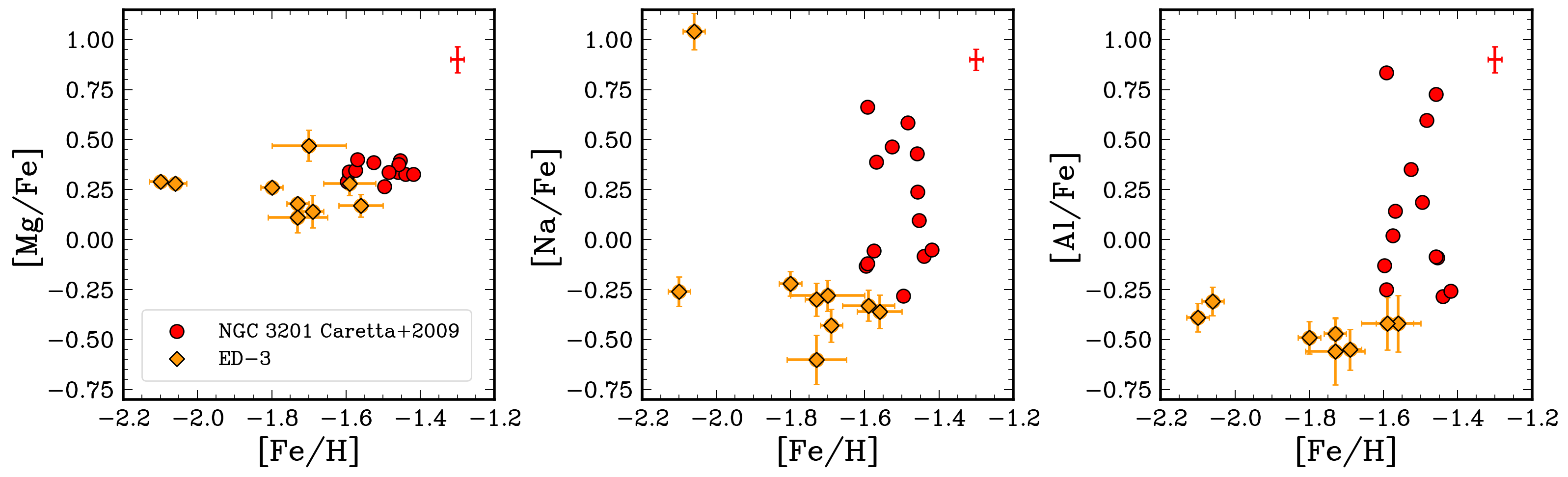}
\caption{ED-3 (orange) light element abundances (\xfe{Mg}, \xfe{Na}, and \xfe{Al}) as a function of \xh{Fe} compared with NGC 3210 abundances from \citet{carretta2009anticorrelation}. The average error bar for NGC 3201 stars are shown in the top-right corner.} 
\label{fig:abundances_3201}
\end{figure*}

If the individual streams in a given ED substructure are chemically consistent, we may then conclude that they belong to the same progenitor and this hints that the progenitor of the substructure is an accreted dwarf rather than a disrupted star cluster, due to the length of the stream. 
We have chemistry for one star in the smaller stream of each ED group, except for ED-4. For ED-3 the star (ED-3-3089) is chemically consistent. In ED-5 the only positive $v_z$ star is ED-5-2653, the high [$\alpha$/Fe] star. For ED-6 the star with positive $v_x$ and chemistry is ED-6-5617 and is chemically consistent.

Another check we can make is if any known GCs could be associated with the ED streams. For this, we use the observables of Galactic GCs given in \citet[][]{vasiliev2021gaia,baumgardt2021accurate} to calculate the IoM of each, sampling within the observable errors.
We show the position of the NGC 3201 in Fig.~\ref{fig:IOM} as a green star, which overlaps with ED-3 in $L_\perp$ and $E$, is only 100 \kms\,kpc off in $L_z$ and lies at a distance of 4.74~kpc from the Sun. None of the other ED-s are on the same orbit as any GC since they do not have the same IoM  ($L_z, L_\perp$ and $E$)~even~when~sampling~within~the~errors.

Before discussing a possible association between ED-3 and NGC 3201 in more detail, we summarise our interpretation of the nature of ED-4. The stars in ED-4 have on average lower $\alpha$ than \textit{Gaia} Enceladus at the same [Fe/H], similar to Sequoia and the Helmi streams (also in other elements). This would suggest a dwarf galaxy origin for ED-4. On the other hand, ED-4 has a relatively small metallicity spread which could be indicative of a star cluster but its stars show no scatter or signs of anti-correlations in Na, Mg or Al abundances. 
Dynamically, a dwarf galaxy origin would be favoured because ED-4 seems to present multiple wraps which would correspond to a larger object. However, this needs to be confirmed with chemical abundances, since all of our analysis comes from stars in the $v_z>0$ clump. In this case, the low metallicity of ED-4 would suggest a low mass dwarf galaxy progenitor \citep[given the mass metallicity relation for galaxies, e.g.][]{kirby2013universal} which is not inconsistent with the relatively small size of ED-4 in IoM space (for example in comparison to the Helmi streams extent in IoM space which are known to stem from a progenitor with a larger mass, e.g. \citealt{koppelman2019a}).

\begin{figure}
\centering
\includegraphics[width=0.5\textwidth]{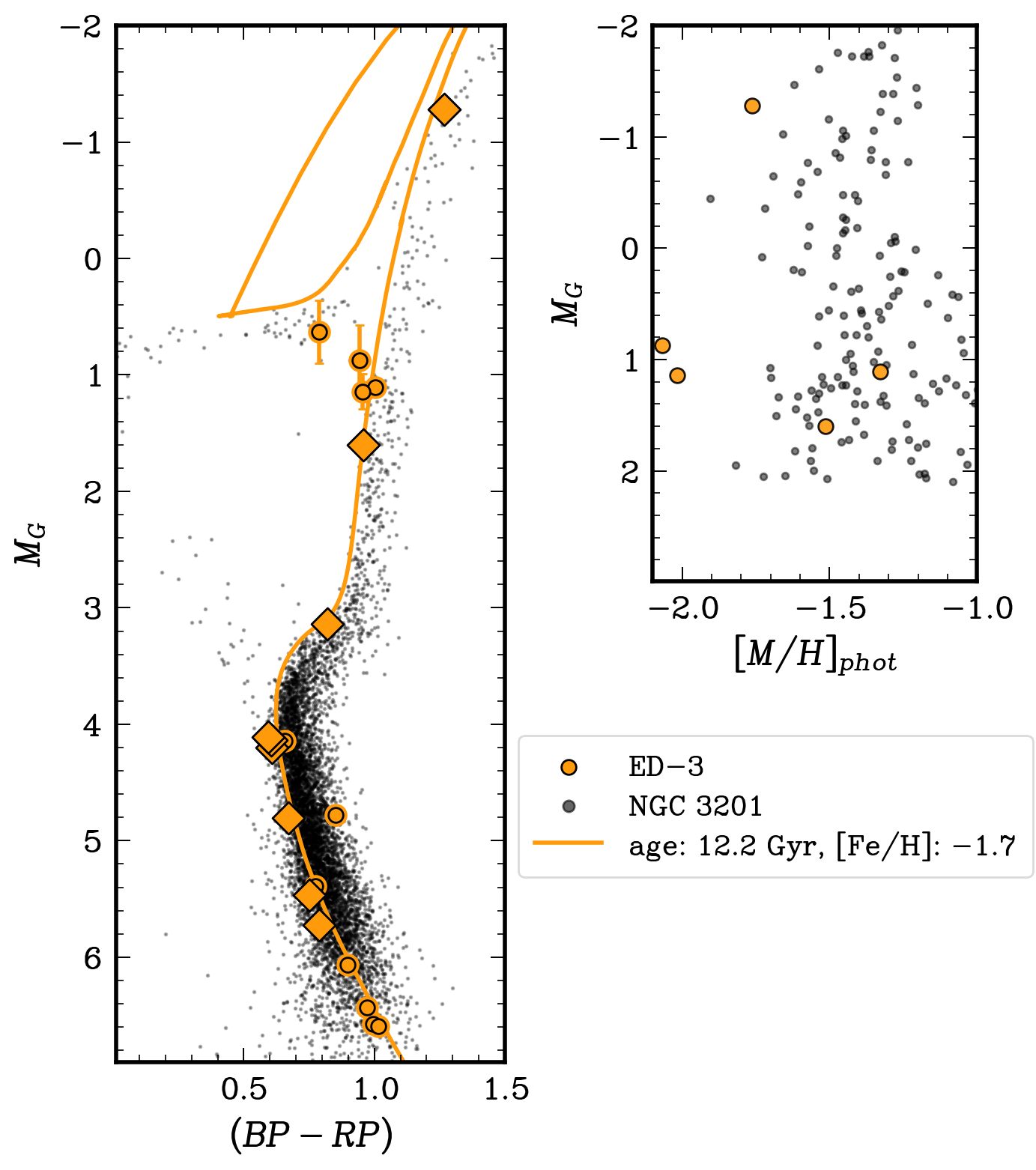}
\caption{Left panel: Colour-magnitude diagram of ED-3 (orange) and NGC 3201 stars (black) with the best-fit isochrone to ED-3 overlaid. While the colours and magnitudes have been corrected for extinction using the 2D dust maps from \citet[][]{schlafly2011measuring}, we caution that there is still a substantial broadening present in the CMD of NGC 3201 due to high differential reddening. Right panel: Photometric metallicities from the \citet{Andrae:2023} quality giant sample as a function of absolute magnitude. A couple of ED-3 stars are similar in their photometric metallicity to NGC 3201. There are two RGB stars in common between our measured \xh{Fe} and \citet{Andrae:2023}, and they are consistent within 0.1dex. } 
\label{fig:CMD_3201}
\end{figure}

\subsection{ED-3 and NGC 3201}\label{sec:ED-3_3201}

Since ED-3 is very close to NGC 3201 in IoM space 
we make a comparison between their stellar populations and chemical abundances. We also integrate the orbit of NGC 3201 and ED-3 using \texttt{AGAMA} \citep{vasiliev2019} in the same potential as described in Sec. \ref{sec:targets}.  NGC 3201's 
orbit brings it within the local 2.5~kpc volume where ED-3 is detected. It is worth noting that a tidal stream from NGC 3201 has been detected previously \citep{palau2021tidal} and the association to a local halo stream Gjöll \citep[][]{ibata2019streams} has been suggested chemically by \citet{hansen2020chemo}. Gjöll is located at a heliocentric distance of $\sim$ 3.4 kpc and the member stars do not overlap with the ED-3 members.

We first compare the detailed elemental abundances derived in this work with those of NGC 3201 presented in \citet{carretta2009anticorrelation}. These are shown in Fig.~\ref{fig:abundances_3201} for \xfe{Mg}, \xfe{Na} and \xfe{Al}. We could see that on average, our abundances for ED-3 are lower in all of \xfe{Mg}, \xfe{Na}, and \xfe{Al} to NGC 3201 stars, but they lie at the tail end of the distributions. There might be systematic offsets in the abundance scales making comparing the absolute [X/Fe] not possible, but the dispersion is more robustly compared. The stars with higher metallicity in ED-3 are more compatible with NGC 3201's metallicity and have a similar spread in \xfe{Mg} to NGC 3201 stars but a significantly smaller spread in \xfe{Al} and \xfe{Na}. 
A spread in light elements suggests the presence of multiple stellar populations that exist in GCs. The fact that we do not observe this in ED-3 makes it difficult to suggest that the high metallicity group is tidal debris from NGC 3201. 

\begin{figure}
\centering
\includegraphics[width=0.5\textwidth]{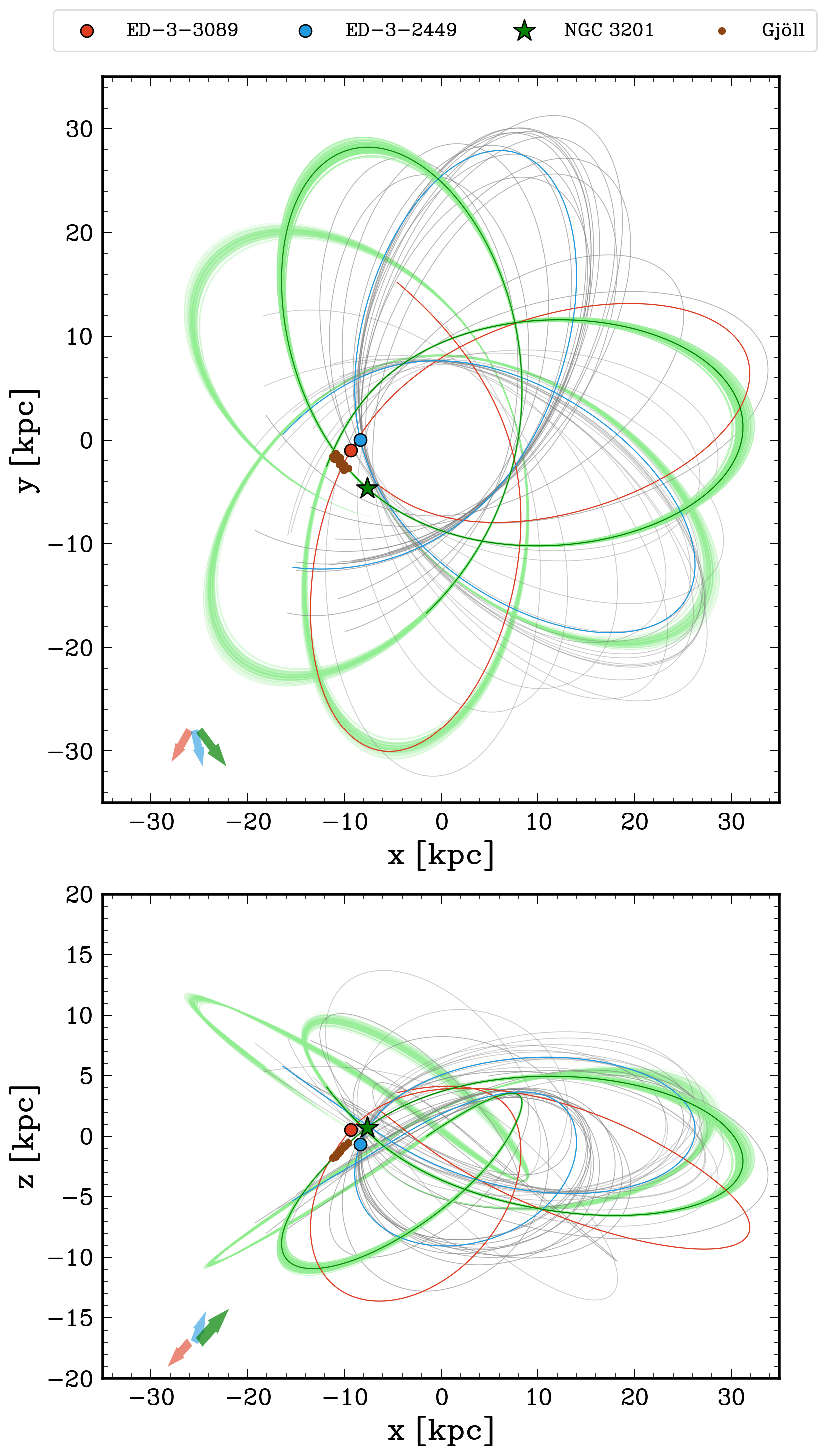}
\caption{Orbits of ED-3 stars and NGC 3201 integrated for 0.5 Gyrs forwards and backwards. Two ED-3 stars are shown with different colours to illustrate the difference in the orbital phase of the stars in $v_z<0$ stream (e.g. ED-3-3089 in red) and the $v_z>0$ stream (e.g. ED-3-2449 in blue). The shaded region around NGC 3201's orbit is 100 samples from the observables within the errors. For NGC 3201 this shading corresponds to a longer integration of 1.5 Gyrs to show how after several revolutions the orbits match the ED-3 stars. Brown points show the Gjöll stream from \textit{Gaia} EDR3 \texttt{STREAMFIDNER} \citep{Ibata:2021}, which overlaps with the trailing part of the orbit of NGC 3201. The arrows in the bottom corner of each panel show the direction of the velocity vector for NGC 3201, ED-3-3089 and ED-3-2449 in green, red and blue respectively. }
\label{fig:orbit_3201}
\end{figure}

We show in Fig.~\ref{fig:CMD_3201} the colour-magnitude diagram of ED-3 stars now with NGC 3201 stars in black for comparison. NGC 3201 stars have been selected using high probability ($>90\%$) members from \citet{vasiliev2021gaia}. 
For the CMD, we made the following quality cuts on the photometry, keeping stars that have a good \texttt{phot\_bp\_rp\_excess\_factor} (see Sect. \ref{sec:stellar_pops}), satisfy the $|C^*| < \sigma_{C^*}$ (where $C^*$ is the corrected BP and RP flux excess from \citealt{riello2021gaia}), and $\beta$<0.1 (BP\_RP blending fraction, which requires a low value to avoid systematics caused by crowding, \citealt{riello2021gaia}). We also required \verb|astrometric_params_solved|=31.
We corrected for extinction using the 2D dust maps from \citet[][a recalibration of \citealt{schlegel1998maps} with the E(B-V) reduced by a factor 0.86]{schlafly2011measuring}, as NGC 3201 is a low latitude cluster for which a single value of E(B-V) is not sufficient, and the \citet{L22} maps underestimate the extinction compared also with \citet[][2010 version]{Harris:1996}. 

We observed in Fig.~\ref{fig:CMD_3201} that there is some overlap between ED-3 stars and NGC 3201 in the stellar populations, but that on average ED-3 stars are bluer. This makes sense given the spread in the metallicities seen in Fig.~\ref{fig:MDFs}. In the second panel of Fig.~\ref{fig:CMD_3201}, the photometric metallicities from the high-quality giants sample of \citet[][Table 2]{Andrae:2023} for ED-3 and NGC 3201 stars can be seen. While the photometric metallicities may have a shift (and show a larger spread) when compared with the spectroscopically derived metallicity\footnote{mean [Fe/H] from \citet{Carretta:2009} is $-1.51$~dex and [Fe/H] from \citet[][2010 version]{Harris:1996} is $-1.59$~dex}, in this case we can compare the two samples directly. We observed that a handful of ED-3 stars have a photometric [M/H] and that some of these overlap with that of NGC 3201 stars.

From the dynamical point of view, we know that ED-3 and NGC 3201 lie on a very similar orbit. We could see this more clearly in Fig.~\ref{fig:orbit_3201} where we show the orbits of ED-3 stars in grey (with one star in each of the $v_z$ streams highlighted) and NGC 3201 in green, integrated for $\sim$ 0.5 Gyrs 
forwards and backwards in time. We show, as shading around NGC 3021's orbit, 100 random orbits with initial conditions sampled from NGC 3021's observables and uncertainties. For this, we show a longer integration of 1.5 Gyrs, forwards and backwards in time. 
We include here also the Gjöll stream from \citet{Ibata:2021}, which matches the trailing arm of NGC 3201. 
From the orbits of NGC 3201 and ED-3, we observed that they do not have the same orbital phase, but their orbits are similar, as expected given their similar IoM.
ED-3 has two clumps in $v_z$ (i.e. stars on different phases of the orbit), and since both contain stars with \xh{Fe} $\sim -1.7$ dex, they do not have a one-to-one correspondence with the two peaks in \xh{Fe} seen in Fig.~\ref{fig:MDFs}. 

\begin{figure*}
\centering
\subfloat[ED-2]{\includegraphics[width = \textwidth]{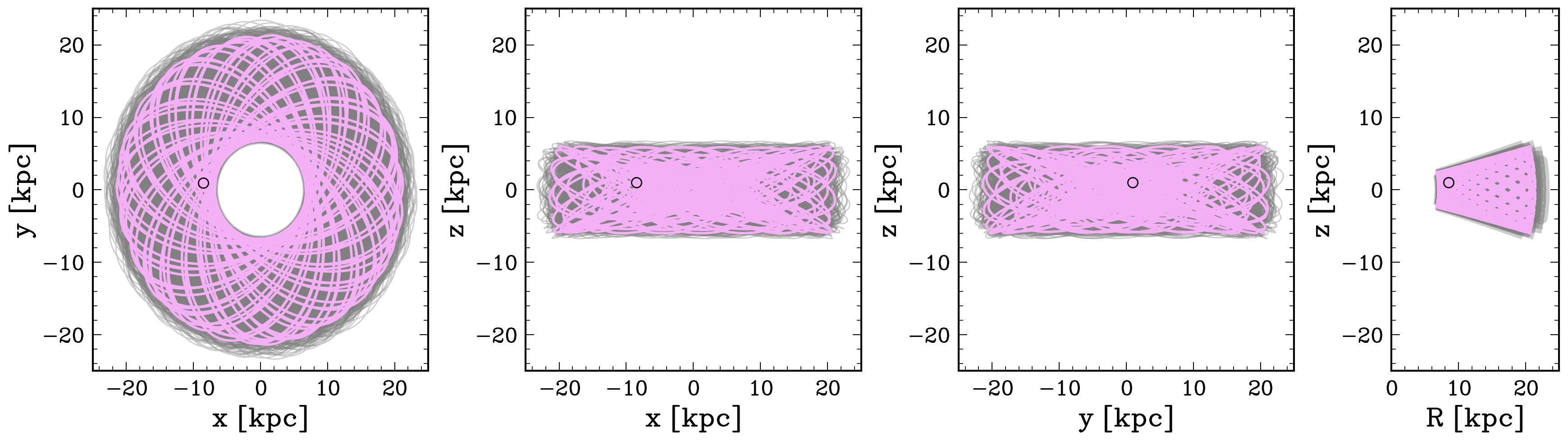}} \\
\subfloat[ED-3]{\includegraphics[width = \textwidth]{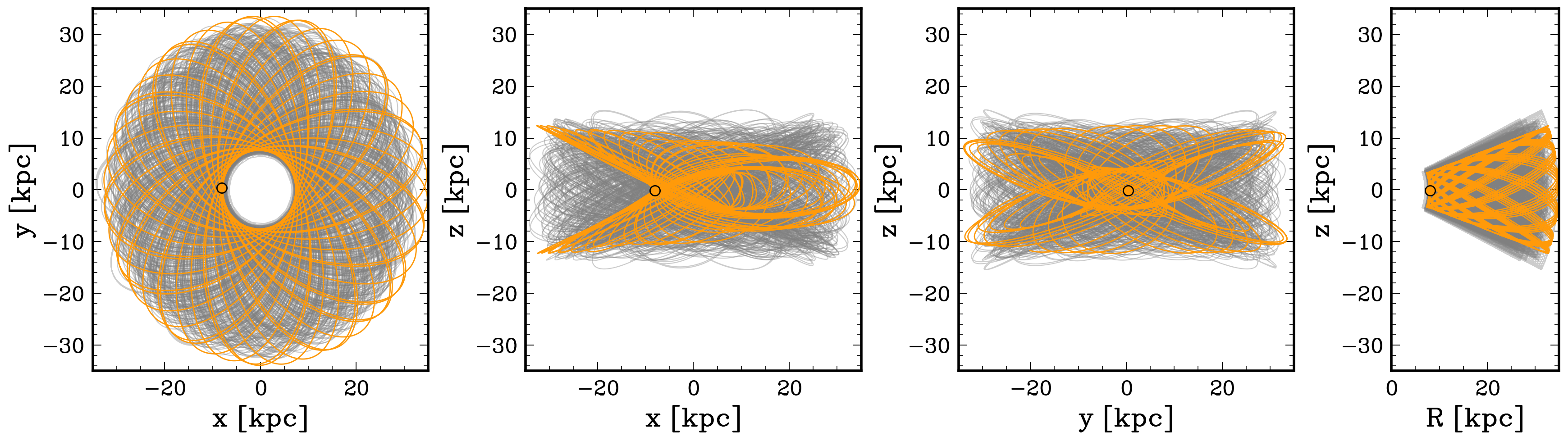}} \\
\subfloat[ED-4]{\includegraphics[width =\textwidth]{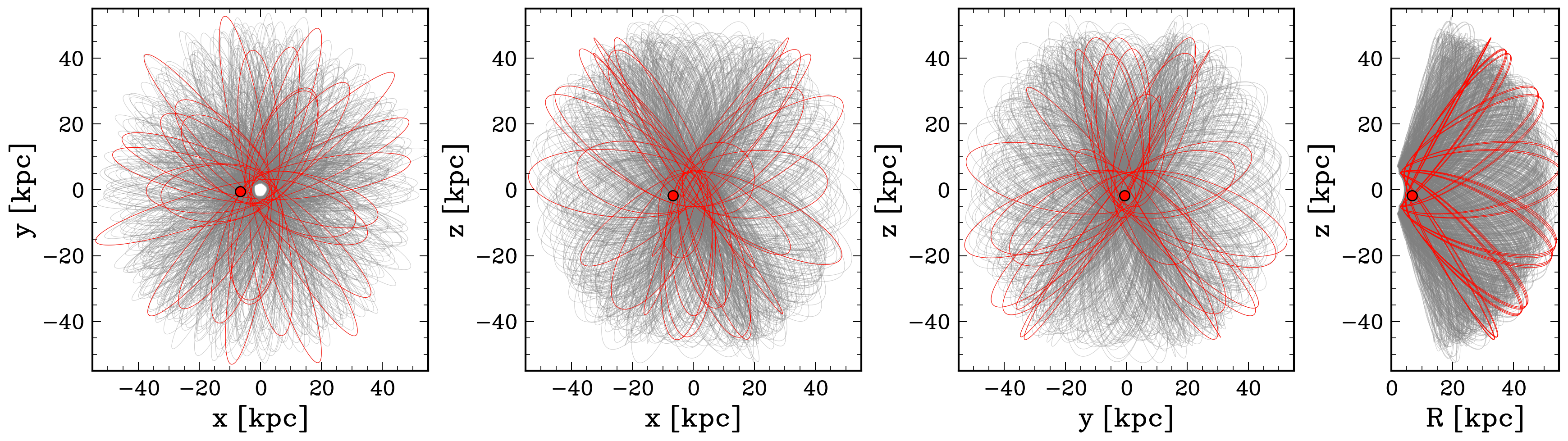}} \\
\caption{Orbit of all stars (grey) in each group integrated for 25 Gyrs and one random star from that group in colour on top. In some cases (e.g. ED-4), the stars exhibit resonant-like orbits.  }
\label{fig:orbits1}
\end{figure*}

\begin{figure*}
\centering
\subfloat[ED-5]{\includegraphics[width =\textwidth]{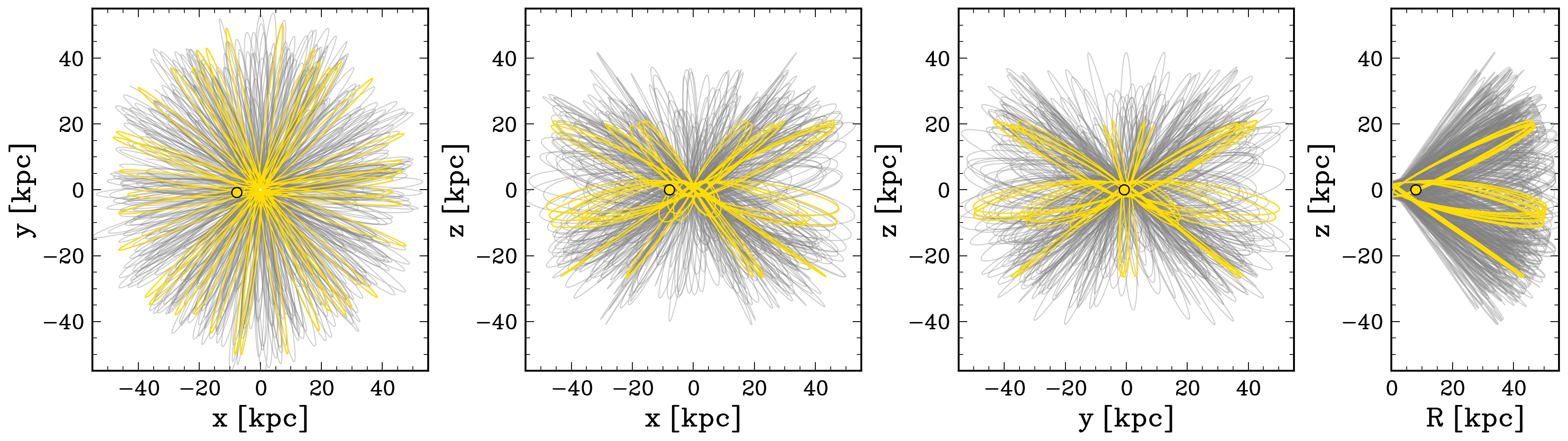}} \\
\subfloat[ED-6]{\includegraphics[width =\textwidth]{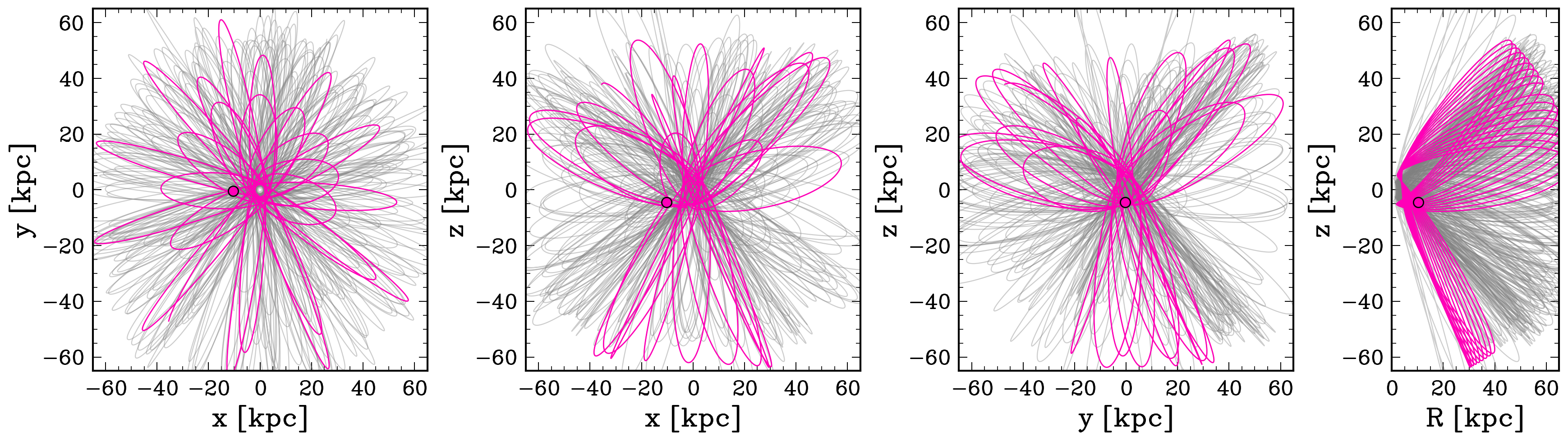}} \\
\caption{Same as Fig.~\ref{fig:orbits1} but for the remaining groups.  }
\label{fig:orbits2}
\end{figure*}

The $v_z>0$ ED-3 stars (e.g. ED-3-2449 shown in blue in Fig.~\ref{fig:orbit_3201}) have close overlap with the expected trailing part of the orbit from NGC 3201, although after three radial revolutions away from NGC 3201. The ED-3 stars with $v_z<0$ (e.g. ED-3-3089 shown in red in Fig.~\ref{fig:orbit_3201}) overlap with the leading part of NGC 3201's orbit, although one radial revolution away. These findings suggest that the stream would be very long. Therefore, while some association to NGC 3201 is clear, it is difficult to explain ED-3 fully as only the tidal debris from the GC NGC 3201, coupled also with ED-3's chemistry (lower $\alpha$ than \textit{Gaia} Enceladus at the same [Fe/H] and a detectable \xh{Fe} spread). Our findings would suggest ED-3 is the debris from a small dwarf galaxy that NGC 3201 is associated with, although detailed models would be required to confirm this interpretation. Given the location in IoM space, it is worth mentioning that ED-3 stars are kinematically distinct from Sequoia and contains no more metal-rich stars, which would be expected if it was a subset of Sequoia.

\subsection{Resonant orbits and tight clumps in IoM space}\label{sec:overview}
As discussed, several ED streams appear to be chemically consistent with an accreted dwarf origin, which may seem in contradiction with their tightness in IoM space. 
To investigate this further, we integrated the orbits of the stars in all ED streams for $\sim$ 25 Gyrs 
in total. 
These are presented in Figs \ref{fig:orbits1} \& \ref{fig:orbits2}, including all stars in each ED group in grey in the background and one star from that group in colour on top to highlight the peculiarities of the orbits. We notice that the orbits of all of ED-4, -5 and -6 (and potentially also ED-3) present stars on orbits that appear very close to resonant. This can be seen by the fact that they do not fill the full volume in position space, for example, they do not fill the available region in $R-z$ space, even after integrating for a sufficiently long time. This is the case even when the choice of Milky Way potential is varied, including the time-dependent potentials from \citet{sormani2022stellar} and \citet{vasiliev2024dear}.
ED-4, -5 and -6 have larger apocenters due to their orbits with lower binding energy and also show signs of being close to an orbital resonance.

The assumption that a tight clump in IoM space stems from a small progenitor (e.g. a small dwarf galaxy or a globular star cluster, with a small spread in orbits) breaks down if the debris falls close to a resonant orbit. When a group of stars is on or close to a resonance, phase mixing occurs on a longer timescale \citep{vogelsberger2008}. 
This makes substructures more detectable and could potentially explain why we detect several of the ED streams and also why they are so tight in IoM despite being seemingly associated with a dwarf galaxy progenitor. Furthermore, the presence of a separatrix around the resonance may result in a chaotic region in which stars will rapidly diffuse away from one another \citep[e.g.][]{price-whelan2016} having fewer integrals of motion, leaving only those on the resonance tightly clumped in IoM space (e.g. as seen for the Helmi streams; \citealt{dodd2022,woudenberg2024first}).
These findings are interesting and could explain the detectability of the ED streams, although this is something that needs to be investigated further.

\section{Conclusion}\label{sec:conclusion}
We have presented detailed chemical abundances of local halo stars in the ED streams, including those measured from the newly obtained UVES spectra of 20 stars and from the archive spectra of nine stars. We coupled these data with homogeneous abundances from archival spectra of 12 GE stars for comparison, and this allowed us to provide new insights into the nature of these streams. 

Our findings suggest that the ED streams are all likely of an accreted origin. All of these systems are on low binding energy or retrograde orbits, and most of the stars analysed in this work have chemical abundances in line with an accreted origin (low \xfe{Na}, \xfe{Al}, and \xfe{\alpha}), except for two stars with a high $\alpha$ at a high \xh{Fe}. 
How these two high-$\alpha$ in situ stars ended up on such low binding energy orbits remains an open and interesting question. Accurate ages of these stars could shed light on the early Milky Way and provide insights as to when and how these stars, likely the proto-disc, were splashed onto these orbits.

Our results confirm recent work showing that the progenitor of ED-2 is likely an ancient star cluster \citep[][]{balbinot2024}. However, we did not detect light element abundance variations that are characteristic of GCs. A homogeneous analysis of more stars is needed to investigate this further. The other streams present a detectable spread in [Fe/H] and multiple clumps in velocity space (multiple wraps of the same stream crossing the solar neighbourhood) that are suggestive of a dwarf galaxy origin. ED-5 and ED-6 present a higher mean metallicity, a larger dispersion, and abundances that are consistent with the \textit{Gaia} Enceladus track (when removing the outlier high-$\alpha$ stars). This suggests that they are likely high-energy tails of GE that are expected to cross the solar neighbourhood, and they could be useful for placing constraints on the configuration of this merger \citep{koppelman2020}. 
It would be interesting to obtain precise ages for the stars in these systems in order to see if this information can be coupled with more tailored modelling of the GE accretion event to constrain further the nature of GE’s progenitor.
ED-3 and ED-4 tend to exhibit a lower [$\alpha$/Fe] compared to the other streams and GE stars, aligning more with the Sequoia and Helmi streams. This suggests a slower chemical enrichment history, with the overall picture suggesting that these are small dwarf galaxies. There is no obvious known accretion event for which ED-3 or ED-4 matches, although our analysis shows that there are strong indications that NGC 3201 is associated with ED-3.

It is intriguing that although all the substructures are very tight in IoM space, they are likely to be the debris of dwarf galaxies (except ED-2). This is probably due to the orbits of these systems being close to resonant. It would be worthwhile to investigate how many of these resonances can be expected in the local halo and their location as well as to analyse how substructures arise and are enhanced as a consequence.

\begin{acknowledgements} 
This research has been partially funded by a Spinoza award by NWO (SPI 78-411). ES acknowledges funding through VIDI grant "Pushing Galactic Archaeology to its limits" (with project number VI.Vidi.193.093) which is funded by the Dutch Research Council (NWO). TRL acknowledges support from Juan de la Cierva fellowship
(IJC2020-043742-I) and Ram\'on y Cajal fellowship (RYC2023-043063-I,
financed by MCIU/AEI/10.13039/501100011033 and by the FSE+).
This work has made use of data from the European Space Agency (ESA) mission
{\it Gaia} (\url{https://www.cosmos.esa.int/gaia}), processed by the {\it Gaia}
Data Processing and Analysis Consortium (DPAC,
\url{https://www.cosmos.esa.int/web/gaia/dpac/consortium}). Funding for the DPAC
has been provided by national institutions, in particular, the institutions
participating in the {\it Gaia} Multilateral Agreement.
The analysis has benefited from the use of the following packages: vaex \citep{breddels2018}, numpy \citep{van2011}, matplotlib \citep{hunter2007}, scipy \citep{2020SciPy-NMeth}, dustmaps \citet{green2018_dustmaps} and jupyter notebooks \citep{kluyver2016}.
\end{acknowledgements}

\bibliographystyle{aa} 
\bibliography{references}

\appendix
\begin{table*}[h!]
\section{Supplementary tables}
	\centering
	\caption{List of stars for which we have measured chemical abundances for in this work.
 }
	\begin{tabular}{lccccccr}
		\hline
		\hline
		\textit{Gaia} ID        & Group      & Name & Obs. Date.     & proposal ID & RV (\kms) \\
		\hline
		\hline
		4245522468554091904$^\textbf{*}$ & ED-2       & ED-2-4245        & 10 Aug 2023  & 111.D-0263A                 
  &-295.70       \\
		\multirow{2}{*}{4479226310758314496}              & \multirow{2}{*}{ED-2}       & \multirow{2}{*}{ED-2-4479}     & 30 May 2022   & 109.B-0522A             
  &-264.07       \\
		                                 &            &                  & 29 Aug 2001   & 167.D-0173A             
                                   &-263.60       \\
		6632335060231088896$^\textbf{*}$ & ED-2       & ED-2-6632        & 24 Aug 2023 & 111.D-0263A               
 & 345.53       \\
		6746114585056265600$^\textbf{*}$ & ED-2       & ED-2-6746        & 10 Aug 2023  & 111.D-0263A               
  & 83.87        \\
		4318465066420528000              & ED-2 \tiny{(BH3 companion)} & ED-2-4318        & 4 Nov 2020   & 106.B-0664A      
  & -383.20            \\
		\hline   
  
		2449962874510227328$^\textbf{*}$ & ED-3       & ED-3-2449        & 28 Aug 2023    & 111.D-0263A           
  &  -178.72   \\
		2551243597828253312$^\textbf{*}$ & ED-3       & ED-3-2551        & 2 Aug 2023     & 111.D-0263A            
  &  -406.81    \\
         2910503176753011840 & ED-3 & ED-3-2910 & 14, 16 Apr 2015&  095.D-0504A 
         & 254.03\\
         
		3089944573918027648              & ED-3       & ED-3-3089        & 16 Nov 2020    & 106.B-0664A            
  & 395.95       \\
		4425854676297423104              & ED-3       & ED-3-4425        & 30 May 2022    & 109.B-0522A            
  &   23.16      \\
		4687536553944071168$^\textbf{*}$ & ED-3       & ED-3-4687        & 24 Aug 2023    & 111.D-0263A           
  &   232.70      \\
		5382632652358260864              & ED-3       & ED-3-5382        & 21 Nov 2019    & 104.B-0487B   
  &   585.56      \\
		6334970766103389824$^\textbf{*}$ & ED-3       & ED-3-6334        & 26 Aug 2023    & 111.D-0263A         
  &   170.55      \\
		6557154165968534400$^\textbf{*}$ & ED-3       & ED-3-6557        & 23 Aug 2023    & 111.D-0263A        
  &    -211.50    \\
		\hline   
  
		2436947439975372672              & ED-4       & ED-4-2436        & 04 Nov 2020 & 106.B-0664A    
  &  -324.61     \\
		4653842020087501824$^\textbf{*}$ & ED-4       & ED-4-4653        & 2 Sep 2023  & 111.D-0263A       
  &  -141.47     \\
		4779027844180362368$^\textbf{*}$ & ED-4       & ED-4-4779        & 2 Sep 2023  & 111.D-0263A            
  &  -43.88      \\
		5012933996503765504$^\textbf{*}$ & ED-4       & ED-4-5012        & 27 Aug 2023 & 111.D-0263A         
  &  -212.95      \\
		5195968563310843008              & ED-4       & ED-4-5195        & 23 Oct 2001 &167.D-0173A            
  &  -95.03      \\
		6434310817041300096$^\textbf{*}$ & ED-4       & ED-4-6434        & 23 Aug 2023 & 111.D-0263A            
  &   -239.28     \\
		\hline   
  
		2653208801494625792$^\textbf{*}$ & ED-5       & ED-5-2653        & 23 Aug 2023  &  111.D-0263A         
  &  -303.91      \\
		3802597128565139584              & ED-5       & ED-5-3802        & 16 May 2022  &109.B-0522A            
  &   -33.88     \\
		5343715644471712256$^\textbf{*}$ & ED-5       & ED-5-5343        & 24 Jul 2023 &  111.D-0263A              
  &   364.36     \\
		6039253574766562176$^\textbf{*}$ & ED-5       & ED-5-6039        & 27 Aug 2023 &  111.D-0263A             
  &   283.54      \\
		\hline   
  
		2503491051919554304$^\textbf{*}$ & ED-6       & ED-6-2503        & 2 Aug 2023   & 111.D-0263A             
  &  361.50       \\
		3266449244243890176$^\textbf{*}$ & ED-6       & ED-6-3266        & 29 Aug 2023 & 111.D-0263A             
  &  408.36        \\
		4958781983684481024$^\textbf{*}$ & ED-6       & ED-6-4958        & 23 Aug 2023 & 111.D-0263A            
  &  306.68     \\
		5617887596219030016$^\textbf{*}$ & ED-6       & ED-6-5617        & 16 Sep 2023 & 111.D-0263A    
  &  34.38       \\
		6643953840115817728$^\textbf{*}$ & ED-6       & ED-6-6643        & 24 Aug 2023 & 111.D-0263A      
  & -147.49           \\
		\hline    
		1159108770069883136              & GE         & GE-1159          & 20 Aug 2022    & 109.B-0480B       
  &  -144.53  \\
		2601354871056014336              & GE         & GE-2601          & 16 Oct 2005    & 076.D-0451A      
  & -75.01  \\
		2602889858007989632              & GE         & GE-2602          & 08 Aug 2022    & 109.B-0522D      
  & -53.196  \\
		3028486001397877120              & GE         & GE-3028          & 14 Apr 2022    & 109.B-0522A      
  & -18.05  \\
		3822808140853060224              & GE         & GE-3822          & 04 Dec 2012    & 090.B-0605A            
  & 120.81  \\
		3933445776843483008              & GE         & GE-3933          & 21 Mar 2011    & 086.D-0871A            
  & -72.05  \\
		5486881507314450816              & GE         & GE-5486          & \tiny{04 Jun, 23 Aug, 09 Sep, 2022}  & 109.B-0480A 
  & 261.36    \\
		6155896330944952576              & GE         & GE-6155          & 12 Apr 2022    & 109.B-0522A             
  & -9.96    \\
		6175345867006532480              & GE         & GE-6175          & 27 May 2022    & 109.B-0522A             
  & -74.98    \\
		6264994650662003712              & GE         & GE-6264          & 06 Aug 2021    & 105.D-0462A              
  & 328.91    \\
		6382699260195307264              & GE         & GE-6382          & 27 May 2022    & 109.B-0522A           
  & 66.43    \\
		6582557248259076096              & GE         & GE-6582          & 23 Aug 2009    & 083.B-0281A         
  & 218.07    \\
    
		\hline
	\end{tabular}
	\label{tab:obs}
    \tablefoot{Stars marked with an asterisk ($^\textbf{*}$) are new UVES spectra we obtained with our own follow-up (111.D-0263A PI Dodd). The rest of the stars are obtained from querying the archive but reanalysed here for direct comparison. We include here the naming convention used in this work which is the group name followed by the first four digits of the \textit{Gaia} ID and we also provide the date of observation and the radial velocity derived from the spectra.}
\end{table*}

\FloatBarrier
\twocolumn

\begin{table*}[h]
   \centering
   \caption{Stellar parameters for all stars.}
   \begin{tabular}{lccccc}
   \hline
   \hline
  \textit{Gaia} ID & Name & \teff\, (K) & \logg \, (dex) & \vt\,\kms & \xh{Fe} (dex)  \\
    \hline
   \hline

4245522468554091904  & ED-2-4245  &   6544    &   4.292   &   1.60 &   -2.47\\
4479226310758314496  & ED-2-4479  &   6223    &   4.540   &   1.36 &   -2.54\\
6632335060231088896  & ED-2-6632  &   5650    &   3.602   &   1.36 &   -2.43\\
6746114585056265600  & ED-2-6746  &   6180    &   4.555   &   1.32 &   -2.43\\
4318465066420528000  & ED-2-4318 \tiny{(BH3)}  &   5445    &   3.037   &   1.60 &   -2.34\\
\hline
2449962874510227328  & ED-3-2449  &   6317    &   4.290   &   1.20 &   -1.73\\
2551243597828253312  & ED-3-2551  &   5626    &   3.674   &   1.20 &   -1.55\\
2910503176753011840  & ED-3-2910  &   6367    &   4.273   &   1.36 &   -2.10\\
3089944573918027648  & ED-3-3089  &   5156    &   2.883   &   1.24 &   -1.54\\
4425854676297423104  & ED-3-4425  &   5699    &   4.990   &   0.75 &   -1.77\\
4687536553944071168  & ED-3-4687  &   6156    &   4.487   &   1.24 &   -1.80\\
5382632652358260864  & ED-3-5382  &   4655    &   1.510   &   1.87 &   -1.71\\
6334970766103389824  & ED-3-6334  &   6417    &   4.274   &   1.44 &   -2.06\\
6557154165968534400  & ED-3-6557  &   5775    &   4.645   &   0.64 &   -1.68\\

\hline
2436947439975372672  & ED-4-2436  &   5379    &   3.362   &   1.36 &   -1.74\\
4779027844180362368  & ED-4-4479     &   6258    &   4.079   &   1.48 &   -1.96\\
4653842020087501824  & ED-4-4653  &   6315    &   4.395   &   1.32 &   -1.98\\
5012933996503765504  & ED-4-5012  &   6265    &   4.385   &   1.28 &   -1.80\\
5195968563310843008  & ED-4-5195  &   5678    &   4.729   &   1.68 &   -2.12\\
6434310817041300096  & ED-4-6434  &   6340    &   4.416   &   1.24 &   -1.96\\

\hline
2653208801494625792  & ED-5-2653  &   5434    &   2.530   &   1.96 &   -1.09\\
3802597128565139584  & ED-5-3802  &   6304    &   4.183   &   1.20 &   -1.76\\
5343715644471712256  & ED-5-5343  &   5333    &   3.437   &   0.96 &   -1.41\\
6039253574766562176  & ED-5-6039  &   5687    &   2.850   &   1.79 &   -1.74\\
\hline
2503491051919554304  & ED-6-2503  &   6297    &   4.263   &   1.24 &   -1.50\\
3266449244243890176  & ED-6-3266  &   4823    &   2.354   &   1.12 &   -0.85\\
4958781983684481024  & ED-6-4958  &   5921    &   3.831   &   1.24 &   -1.67\\
5617887596219030016  & ED-6-5617  &   5796    &   3.436   &   1.40 &   -1.59\\
6643953840115817728  & ED-6-6643  &   6006    &   4.252   &   0.72 &   -0.95\\
\hline
1159108770069883136  & GE-1159       &   5383    &   4.628   &   1.52 &   -0.96\\
2601354871056014336  & GE-2601       &   5812    &   3.822   &   1.36 &   -1.93\\
2602889858007989632  & GE-2602       &   6232    &   4.105   &   1.40 &   -1.70\\
3028486001397877120  & GE-3028       &   6407    &   4.395   &   1.40 &   -1.33\\
3822808140853060224  & GE-3822       &   5964    &   4.059   &   1.08 &   -1.07\\
3933445776843483008  & GE-3933       &   6244    &   4.484   &   1.32 &   -2.32\\
5486881507314450816  & GE-5486       &   6002    &   4.345   &   1.20 &   -0.98\\
6155896330944952576  & GE-6155       &   6108    &   4.507   &   1.32 &   -2.34\\
6175345867006532480  & GE-6175       &   5865    &   4.569   &   0.52 &   -1.43\\
6264994650662003712  & GE-6264       &   6142    &   4.414   &   1.28 &   -2.16\\
6382699260195307264  & GE-6382       &   6345    &   4.085   &   1.44 &   -1.31\\
6582557248259076096  & GE-6582       &   6228    &   4.159   &   1.28 &   -1.66\\
\hline
   \end{tabular}    
   \label{tab:stellar_params}
\end{table*}

\FloatBarrier
\twocolumn

\section{Corner plots for metallicity spread}\label{sec:appendix_corner_plots}

\begin{figure}[h]
\centering
\includegraphics[width=0.35\textwidth]{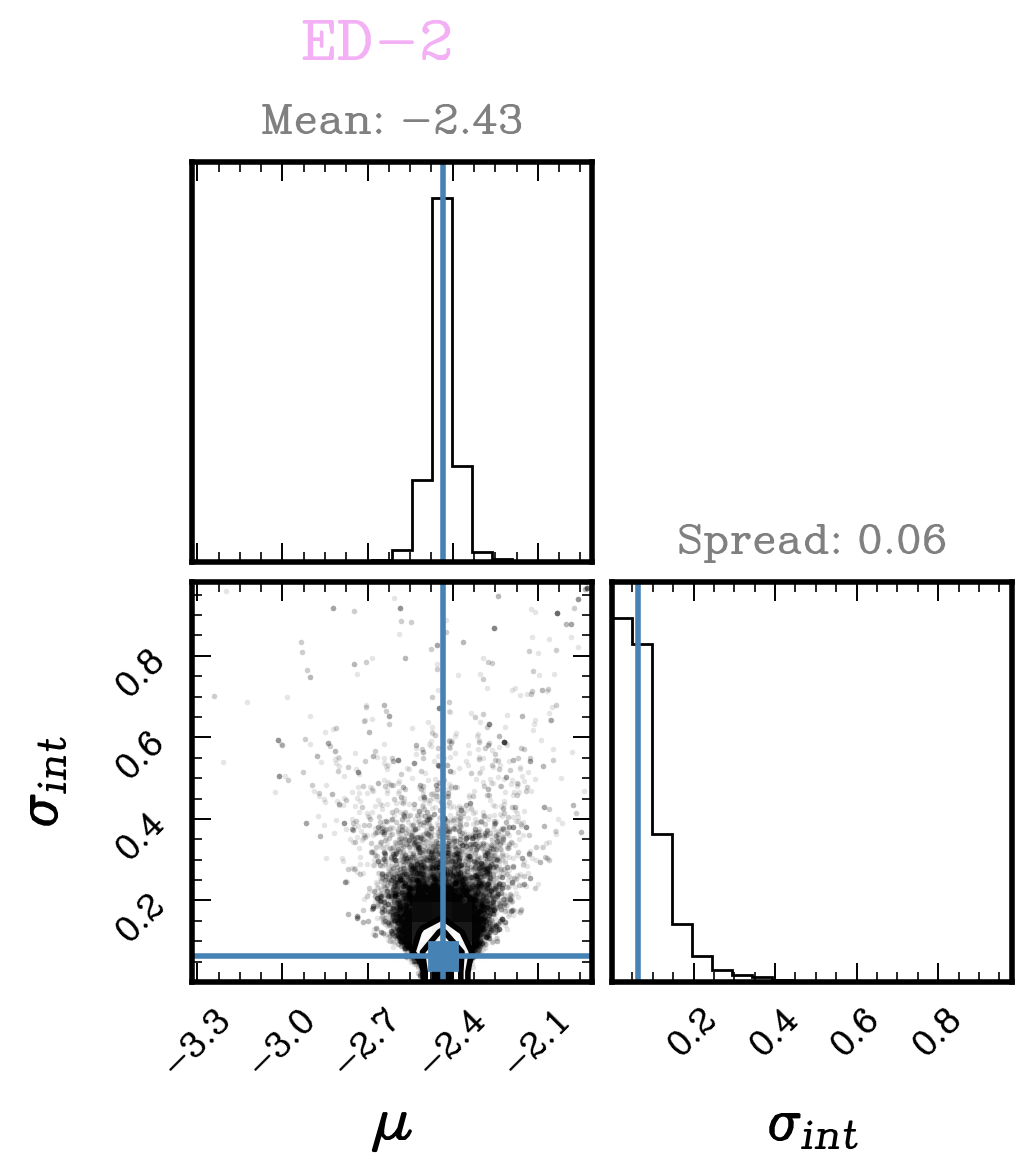}
\includegraphics[width=0.35\textwidth]{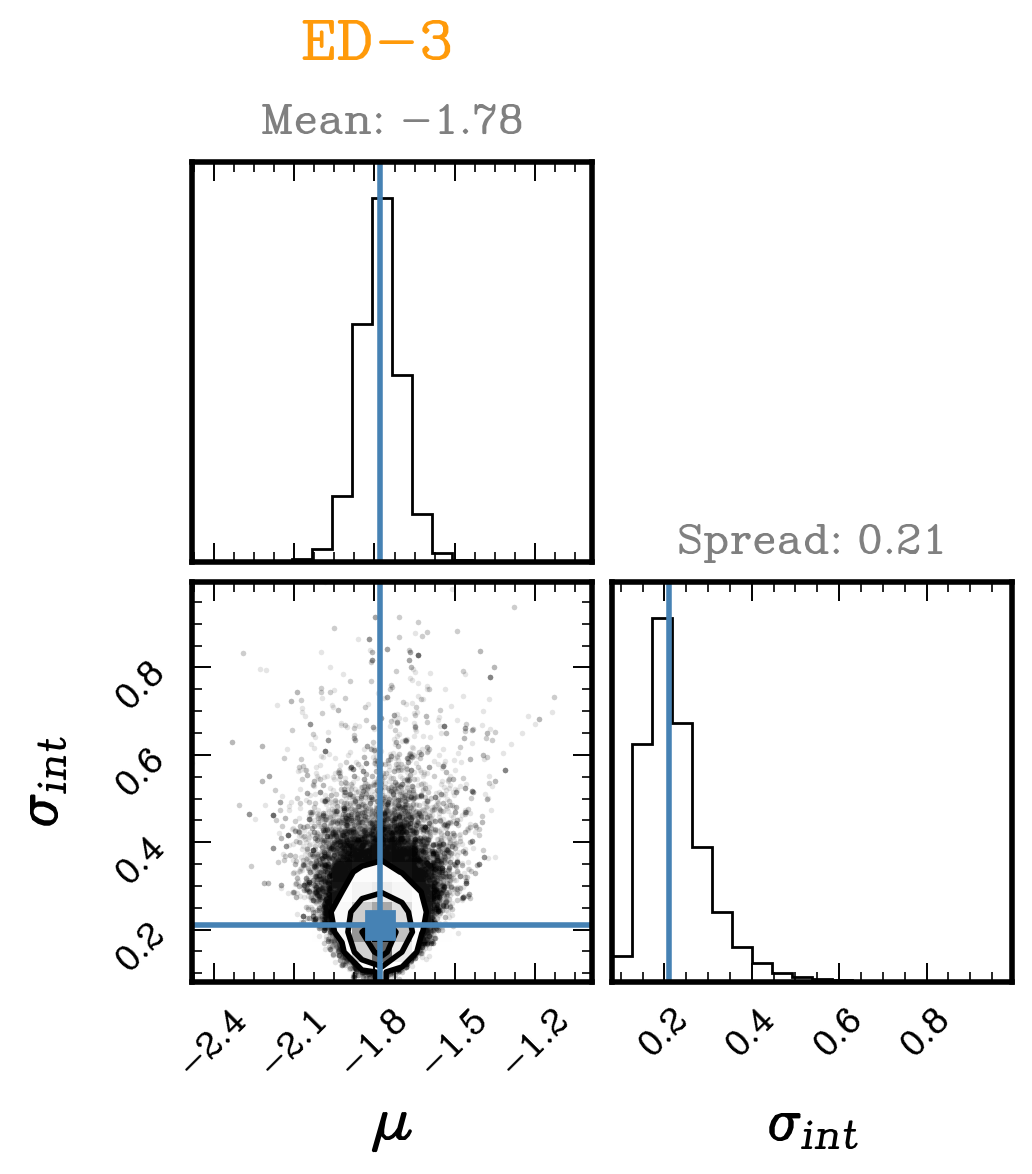}
\includegraphics[width=0.35\textwidth]{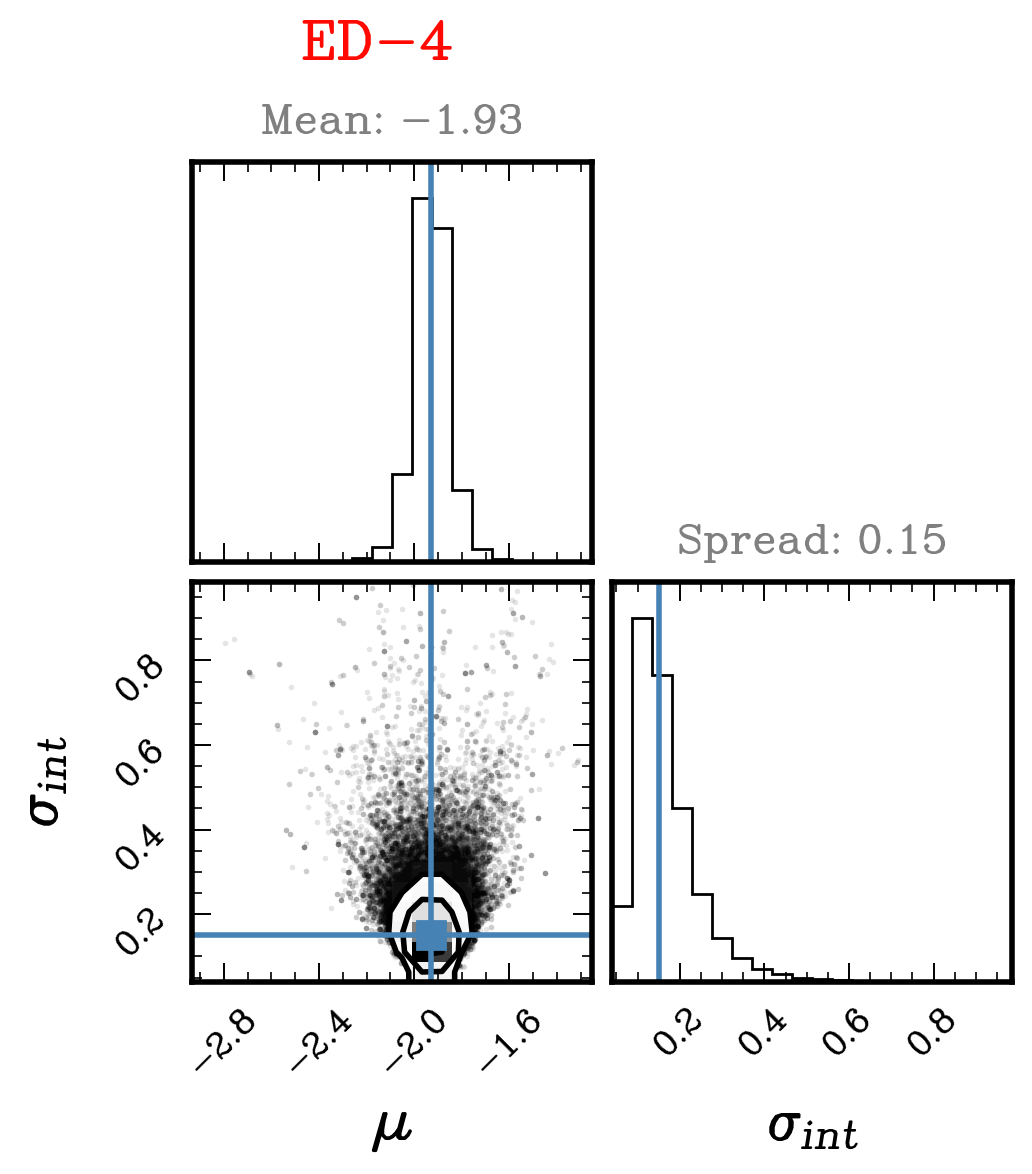}
\caption{MCMC sampling of fitting the mean and intrinsic dispersion of the metallicities of ED-2, ED-3, and ED-4.} 
\label{Fig:corner_plot_ED-2}
\end{figure}

\begin{figure}[h]
\centering
\includegraphics[width=0.35\textwidth]{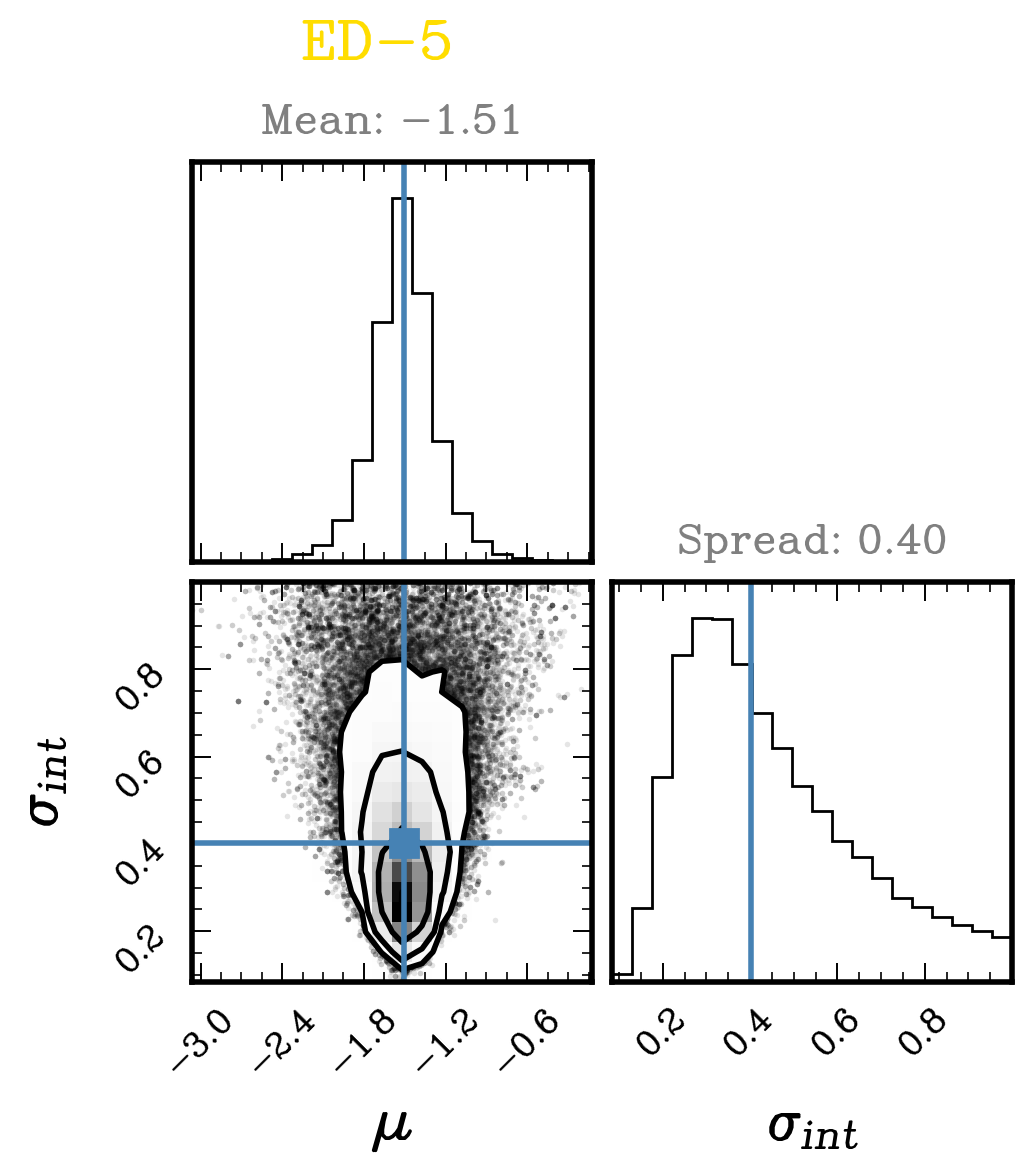}
\includegraphics[width=0.35\textwidth]{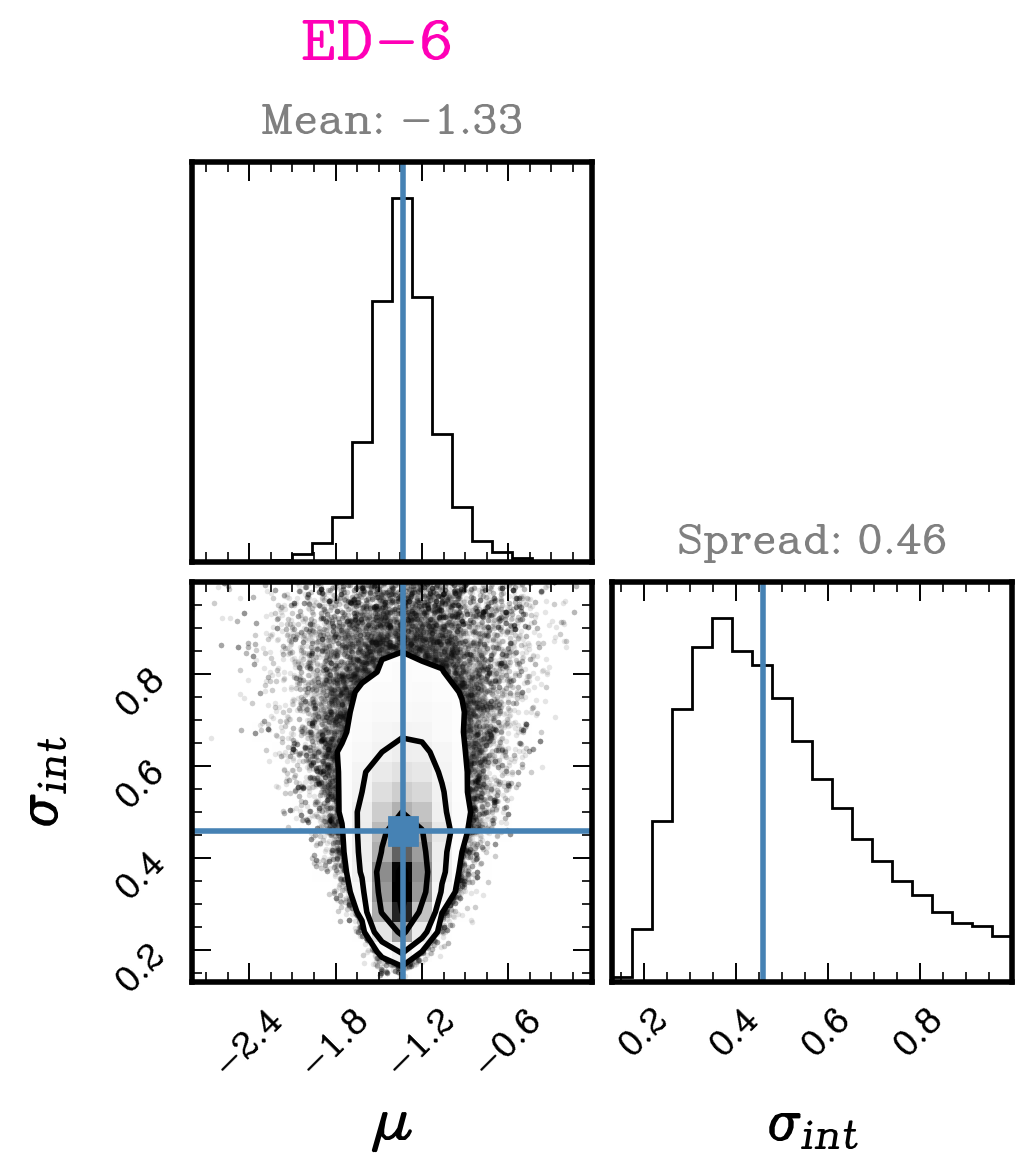}
\caption{Same as Fig. \ref{Fig:corner_plot_ED-2} but for ED-5 and ED-6.} 
\label{Fig:corner_plot_ED-5-6}
\end{figure}

\end{document}